\documentclass[twocolumn,prb,superscriptaddress,longbibliography]{revtex4-1}
\usepackage{graphicx}
\usepackage{dcolumn}
\usepackage{bm}
\usepackage{subfigure}
\usepackage{amsmath}
\usepackage{amssymb}
\usepackage{setspace}
\newcommand{\ket}[1]{|#1\rangle}                      % Ket Dirac's notation %
\newcommand{\bra}[1]{\langle #1|}                     % Bra Dirac's notation  %
\begin{document}
\title{Photonic quantum walk in a single beam with twisted light}
\author{Filippo Cardano}
\affiliation{Dipartimento di Fisica, Universit\`{a} di Napoli Federico II, Complesso Universitario di Monte Sant'Angelo, Napoli, Italy}
\author{Francesco Massa}
\altaffiliation[Current address: ]{Faculty of Physics, University of Vienna, Boltzmanngasse 5, 1090, Vienna, Austria}
\affiliation{Dipartimento di Fisica, Universit\`{a} di Napoli Federico II, Complesso Universitario di Monte Sant'Angelo, Napoli, Italy}
\author{Ebrahim Karimi}
\affiliation{Department of Physics, University of Ottawa, 150 Louis Pasteur, Ottawa, Ontario, K1N 6N5 Canada}
\author{Sergei Slussarenko}
\altaffiliation[Current address: ]{Centre for Quantum Dynamics, Griffith University, Brisbane 4111, Australia}
\affiliation{Dipartimento di Fisica, Universit\`{a} di Napoli Federico II, Complesso Universitario di Monte Sant'Angelo, Napoli, Italy}
\author{Domenico Paparo}
\affiliation{CNR-SPIN, Complesso Universitario di Monte Sant'Angelo, Napoli, Italy}
\author{Corrado de Lisio}
\affiliation{Dipartimento di Fisica, Universit\`{a} di Napoli Federico II, Complesso Universitario di Monte Sant'Angelo, Napoli, Italy}
\affiliation{CNR-SPIN, Complesso Universitario di Monte Sant'Angelo, Napoli, Italy}
\author{Fabio Sciarrino}
\affiliation{Dipartimento di Fisica, Sapienza Universit\`{a} di Roma, Roma 00185, Italy}
\author{Enrico Santamato}
\affiliation{Dipartimento di Fisica, Universit\`{a} di Napoli Federico II, Complesso Universitario di Monte Sant'Angelo, Napoli, Italy}
\author{Lorenzo Marrucci}
\email[Correspondence to: ]{lorenzo.marrucci@unina.it}
\affiliation{Dipartimento di Fisica, Universit\`{a} di Napoli Federico II, Complesso Universitario di Monte Sant'Angelo, Napoli, Italy}
\affiliation{CNR-SPIN, Complesso Universitario di Monte Sant'Angelo, Napoli, Italy}
\begin{abstract}
Inspired by the classical phenomenon of random walk, the concept of quantum walk\cite{Kemp03_ConPhys} has emerged recently as a powerful platform for the dynamical simulation of complex quantum systems\cite{Mohs08_JCP,Kita12_NatCom,Cres13_NatPhot}, entanglement production\cite{Abal06_PRA,Viei13_PRL} and universal quantum computation\cite{Chil09_PRL,Love10_PRA}. Such a wide perspective motivates a renewing search for efficient, scalable and stable implementations of this quantum process. Photonic approaches have hitherto mainly focused on multi-path schemes, requiring interferometric stability and a number of optical elements that scales quadratically with the number of steps \cite{Walt12_NatPhot}. Here we report the experimental realization of a quantum walk taking place in the orbital angular momentum space of light, both for a single photon and for two simultaneous indistinguishable photons. The whole process develops in a single light beam, with no need of interferometers, and requires optical resources scaling linearly with the number of steps. Our demonstration introduces a novel versatile photonic platform for implementing quantum simulations, based on exploiting the transverse modes of a single light beam as quantum degrees of freedom.
\end{abstract}
\maketitle
First proposed by Feynman about thirty years ago\cite{Feyn82}, the simulation of a complex quantum system by means of another well controlled quantum system is nowadays becoming a feasible, although still challenging task. Photons are a reliable resource in this arena, as witnessed by the large variety of photonic architectures that have been realized hitherto for the realization of quantum simulators\cite{Walt12_NatPhot}. Among simulated processes, the quantum walk\cite{Kemp03_ConPhys} (QW) is receiving a wide interest. A QW can be interpreted as the quantum counterpart of the well known classical random walk. In its simplest, one-dimensional (1D) example, the latter is a path consisting of a sequence of random steps along a line. At each step, the walker moves forward or backward according to the outcome of a random process, such as the flip of a coin. When both the walker and the coin are quantum systems we obtain a QW. The final probability distribution for the walker position shows striking differences with respect to the classical process, due to interferences between coherent superpositions of different paths \cite{Knig06_PRA}. It has been demonstrated that this quantum process can be used to perform quantum search algorithms on a graph\cite{Shen03_PRA,Poto09_PRA} and universal quantum computation\cite{Chil09_PRL,Love10_PRA}. Interestingly, it represents a versatile platform for the simulation of phenomena characterizing complex systems, such as Anderson localization in disordered media \cite{Cres13_NatPhot}, topological phases \cite{Kita12_NatCom}, and energy transport in chemical processes \cite{Mohs08_JCP}. In the last decade, implementations of QWs in 1D have been realized in a variety of physical systems, such as trapped ions\cite{Schm09_PRL,Zahr10_PRL} or atoms\cite{Kars09_Sci}, NMR systems \cite{Ryan05_PRA}, and photons, using both bulk optics\cite{Zhan07_PRA,Broo10_PRL,Schr10_PRL} and integrated waveguides\cite{Peru10_Sci,Sans12_PRL,Owen11_NJP}. Remarkably, only few photonic simulations of multi-particles QWs have been reported, using two-photon states\cite{Cres13_NatPhot,Peru10_Sci,Sans12_PRL,Owen11_NJP} or a classical coherent source\cite{Schr12_Sci}. In photonic architectures, different strategies can be adopted, according to the optical degrees of freedom exploited to encode the coin and the walker quantum systems. In 2010 Zhang \textit{et al.} proposed a novel approach for the realization of a photonic walk, based on the idea of encoding the coin and the walker in the spin angular momentum (SAM) and in the orbital angular momentum (OAM) of light, respectively\cite{Zhan10_PRA}. A possible implementation of the same idea in a loop-based configuration has been also analyzed\cite{Goya13_PRL}. These theoretical proposals put forward for the first time the possibility of implementing a photonic walk without interferometers, with the whole process taking place within a single light beam. To obtain this result, both these schemes rely on the spin-orbit coupling occurring in a special optical element called q-plate\cite{Marr06_PRL}, whose action will be discussed later on. In the present work, we implement experimentally the proposal by Zhang \textit{et al.}, thus demonstrating the first photonic QW occurring in a single light beam and using the OAM degree of freedom of photons as discrete walker coordinate. Moreover, we generalize the QW process by introducing in our experiment an adjustable parameter that controls the photon ``mobility'' in the OAM lattice. Finally, in the same platform, we demonstrate the simultaneous QW of two indistinguishable photons propagating in the same beam, thus proving that the method can be extended to higher-dimensional multiparticle systems.
%
%Naive illustration of the process%
\begin{figure}[t]
\centering
\includegraphics[width=8.5cm]{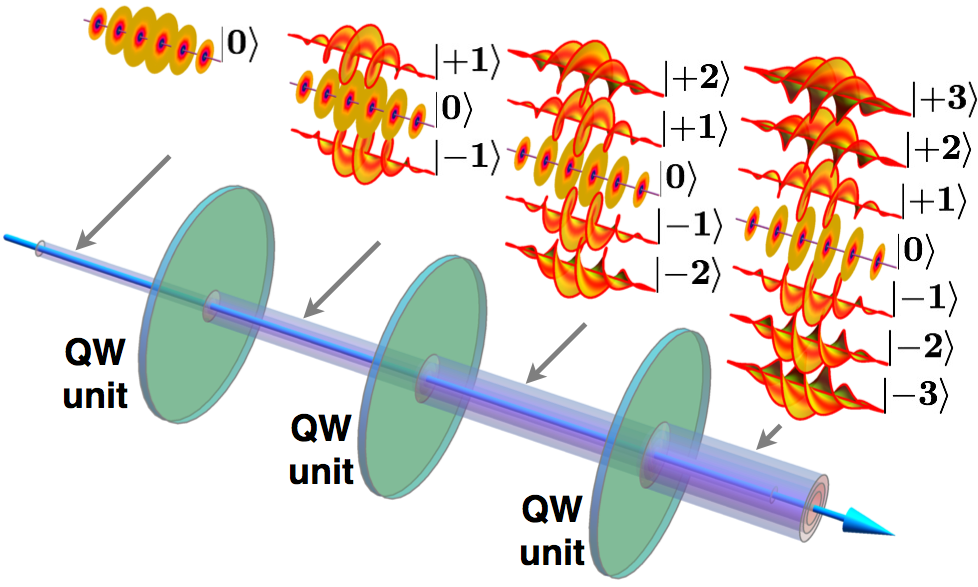}
\caption{Conceptual scheme of the single-beam photonic quantum walk in the space of OAM. In each crossed optical stage (QW unit), the photon can move to an OAM value $m$ that is increased or decreased by one unit (or stay still, in the hybrid configuration). Being this a quantum evolution, the OAM decomposition of the photonic wavefunction at each stage includes many different values, as shown in the call-outs in which modes having different OAM values are represented by the corresponding helical (or ``twisted'') wavefronts.}
\label{fig:naive}
\end{figure}

%%
%
%Experimental layout%
%
\begin{figure*}[t]
\centering
\includegraphics[width=16cm]{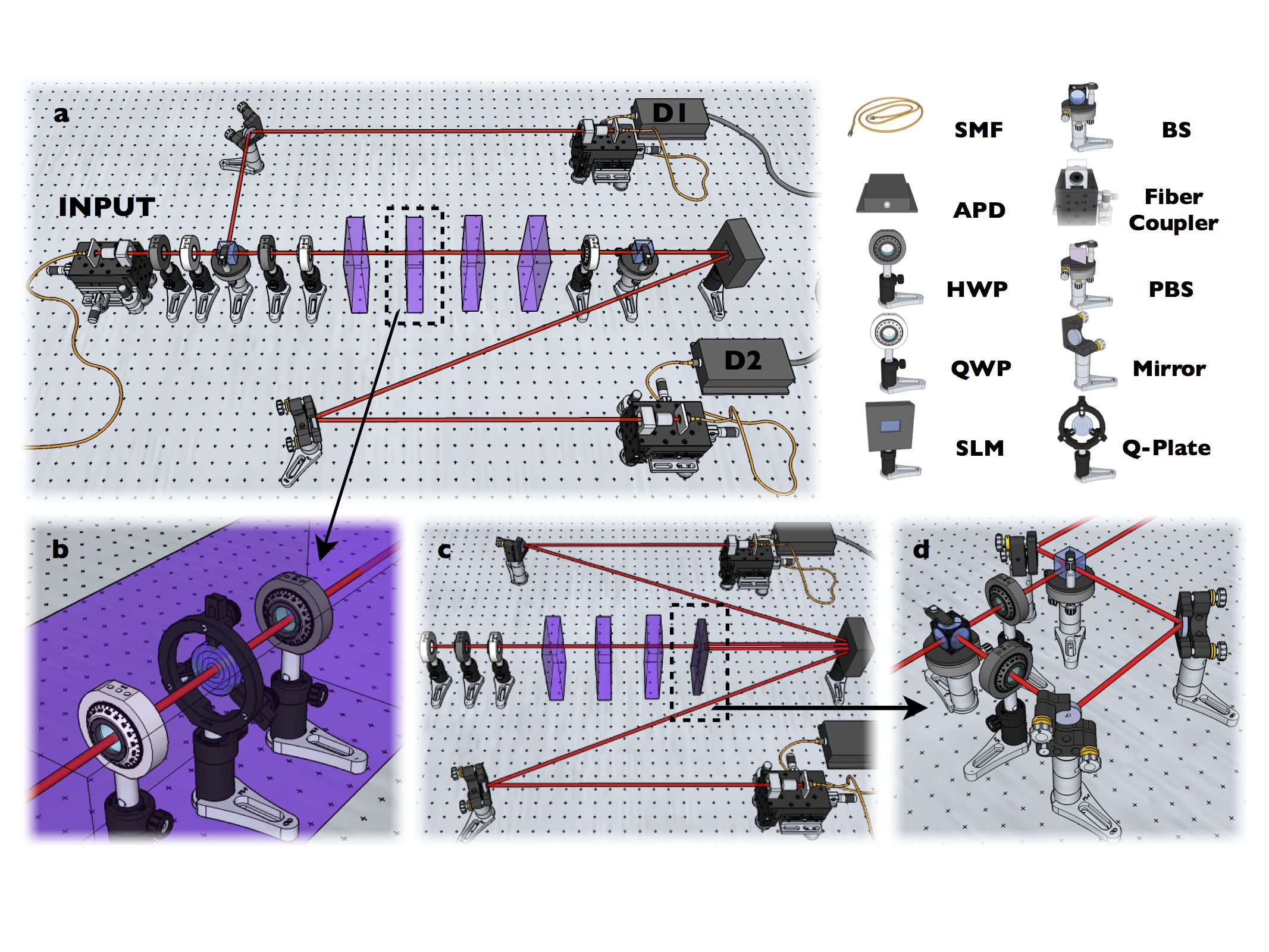}
\caption{Experimental apparatus. a) Setup configuration for single-photon QW experiments. Photon pairs entering the setup are split by a PBS. One photon is sent directly to the avalanche single photon detector (APD) D$_1$, while the other goes through four identical subsequent QW steps (see panel b for the detail of each step) and is then analyzed in polarization and OAM and finally detected with APD D$_2$, in coincidence with D$_1$. The polarization measurement is performed with a HWP-QWP set followed by a PBS. The OAM state is analyzed by diffraction on a spatial light modulator (SLM) followed by coupling into a single-mode optical fiber (SMF). The projection state corresponding to each OAM eigenvalue $m$ was thus fixed by the hologram pattern visualized on the SLM. b)  Detail of the single QW step. The sequence of optical elements in each step includes a QWP with optical axis oriented at 45$^{\circ}$ from the horizontal axis, a q-plate with $q=1/2$, and a HWP with optical axis parallel to the horizontal one. The reference axes of the q-plates are all oriented horizontal. c) Setup configuration for two-photon QW experiments. Both photons exiting the input SMF, after setting their polarization to $L$ and $R$, are sent into a three-steps QW apparatus. The individual QW step is again that showed in panel b). At the QW exit, the two photons are split in two parallel beams (see panel d) in order to measure their joint polarization and OAM probabilities. The polarization-OAM analysis is performed as in the single-photon case, except that two different areas of the SLM are used for diffracting the two photons, which are then coupled into two SMFs and detected in coincidence in APDs D$_1$ and D$_2$ (within a 8 ns time window). d) Detail of the splitting stage. To preserve the polarization state, the splitting is performed on a non-polarizing beam splitter (BS) and relies on post-selection to single-out the cases in which the two photons emerge in different BS ports.}
\label{fig:layout}
\end{figure*}

In the quantum theory framework, a typical QW involves a system described by a Hilbert space $\mathcal H$ obtained by the direct product $\mathcal H_c\otimes\mathcal H_w$ of the coin and the walker subspaces, respectively. In the simplest case, the walker is moving on a 1D lattice and, at each step, has only two different choices. Accordingly, the subspace $\mathcal H_c$ is two-dimensional (2D), while $\mathcal H_w$ is infinite-dimensional; they are spanned by the vectors $\{\ket{\uparrow},\ket{\downarrow}\}$ and $\{\ket{x},\; x\in \mathbb{Z}\}$, respectively. The displacement of the walker at each step of the process is realized by the shift operator $\hat S$
\begin{equation}\label{main:shift}
\hat S=\ket{\uparrow}\bra{\uparrow}\otimes\hat L^++\ket{\downarrow}\bra{\downarrow}\otimes\hat L^-,
\end{equation} 
where the operators $\hat L^{\pm}$ shift the position of the walker, i.e.\ $\hat L^{\pm}\ket{x}=\ket{x\pm1}$. The displacement introduced by $\hat S$ is conditioned by the coin; when this is in the state $\ket{\uparrow}$, the walker moves up, or vice versa. As a consequence, the operator $\hat S$ entangles the coin and the walker systems\cite{Abal06_PRA,Viei13_PRL}. Between consecutive displacements, the ``randomness'' is introduced by a unitary operator $\hat T$ acting on the coin subspace. Usually, $\hat T$ is the Hadamard gate
\begin{eqnarray}\label{main:hada}
\hat T\,\ket{\uparrow}&=\frac{1}{\sqrt 2}(\ket{\uparrow}+\ket{\downarrow})\nonumber\\
\hat T\,\ket{\downarrow}&=\frac{1}{\sqrt 2}(\ket{\uparrow}-\ket{\downarrow}).
\end{eqnarray} 
A single step of the walk is described by the operator $\hat U=\hat S\cdot(\hat T\otimes \hat I_w)$, where $\hat I_w$ is the identity operator in $\mathcal H_w$. After $n$ steps, the system initially prepared in the state $\ket{\psi_0}$ evolves to a new state
\begin{equation}
\ket{\psi_n}= \hat U^n\ket{\psi_0}.
\end{equation} 
Consider now a photon and its internal degrees of freedom represented by the SAM and the OAM. In the limit of paraxial optics, these two quantities are independent and well defined; the first is associated with the polarization of the light, while the second is related to the azimuthal structure of the photonic wave function in the transverse plane \cite{Marr11_JOpt}. The SAM space $\mathcal H_p$ is spanned by vectors $\{\ket{L},\ket{R}\}$, representing left-circular and right-circular polarizations. The OAM space $\mathcal H_o$ is spanned by vectors $\ket{m}$ with $m\in \mathbb{Z}$, which denote a photon carrying $m\hbar$ of OAM along the propagation axis and having a correspondingly ``twisted'' wavefunction (see Fig.\ \ref{fig:naive}).

In our implementation, the coin and the walker systems are encoded in the SAM and the OAM of a photon, respectively. In particular, the spatial walker coordinate $x$ is replaced by the OAM coordinate $m$. The concept of a QW in OAM within a single optical beam is pictorially illustrated in Fig.\ \ref{fig:naive}. The step operator $\hat U$ is implemented by means of linear-optical elements. In the coin subspace, the Hadamard gate is simply a quarter-wave plate (QWP). The shift operator $\hat S$ is instead realized by a q-plate (QP), a recently-introduced photonic device which has already found many useful applications in classical and quantum optics \cite{Marr06_PRL,Marr11_JOpt}. The QP is a birefringent liquid-crystal medium with an inhomogeneous optical axis that has been arranged in a singular pattern, with topological charge $q$, so as to give rise to an engineered spin-orbit coupling in the light crossing it. In particular, the QP raises or lowers the OAM of the incoming photon according to its SAM state, while leaving the photon in the same optical beam, i.e.\ with no deflections nor diffractions. In the actual device also the radial profile of the photonic wave function undergoes a small alteration (as long as it remains in the near-field regime), which however can be approximately neglected in our implementation, as discussed in the Supplementary Information (SI). More precisely, the action of a QP can be generally described by the operator $\hat Q_\delta$
\begin{eqnarray}\label{main:qplate}
\hat Q_\delta \ket{L,m} &= \cos{(\delta/2)}\ket{L,m}-i\sin{(\delta/2)}\ket{R,m+2q} \nonumber\\
\hat Q_\delta\ket{R,m}&= \cos{(\delta/2)}\ket{R,m}-i\sin{(\delta/2)}\ket{L,m-2q}, 
\end{eqnarray}
where $q$ is the QP topological charge  and $\delta$ the optical birefringent phase-retardation \cite{Marr06_PRL,Marr11_JOpt}. While $q$ is a fixed property of the q-plate, $\delta$ can be controlled dynamically by tuning an applied voltage \cite{Picc10_APL}. As shown in Eq.\ \ref{main:qplate}, the action of the q-plate is made of two terms. The first, proportional to $\cos(\delta/2)$, leaves the photon in its input state. The second, proportional to $\sin(\delta/2)$, implements the conditional displacement of Eq.\ \ref{main:shift}, but adds also a flip of the coin state. The latter effect can be however compensated by inserting an additional half-wave plate (HWP). When $\delta=\pi$ (``standard'' configuration) the first term vanishes and the standard shift operator $\hat S$ is obtained. When $\delta=0$, the evolution is trivial (the walker stands still), while for intermediate values $0<\delta<\pi$ we have a novel kind of evolution: besides moving forward or backward, the walker at each step is provided with a third option, that is to remain in the same position. We refer to this as ``hybrid'' configuration, since it mimics a walk with three possible choices, although the coin is still two-dimensional. Similar to an effective mass, the $\delta$ parameter controls the degree of mobility of the walker, ranging from a vanishing mobility for $\delta=0$ to a maximal mobility (such as that occurring for massless particles) for $\delta=\pi$. 
%
%Data 1 photon%
%
\begin{figure*}[t]
\centering
\includegraphics[width=16cm]{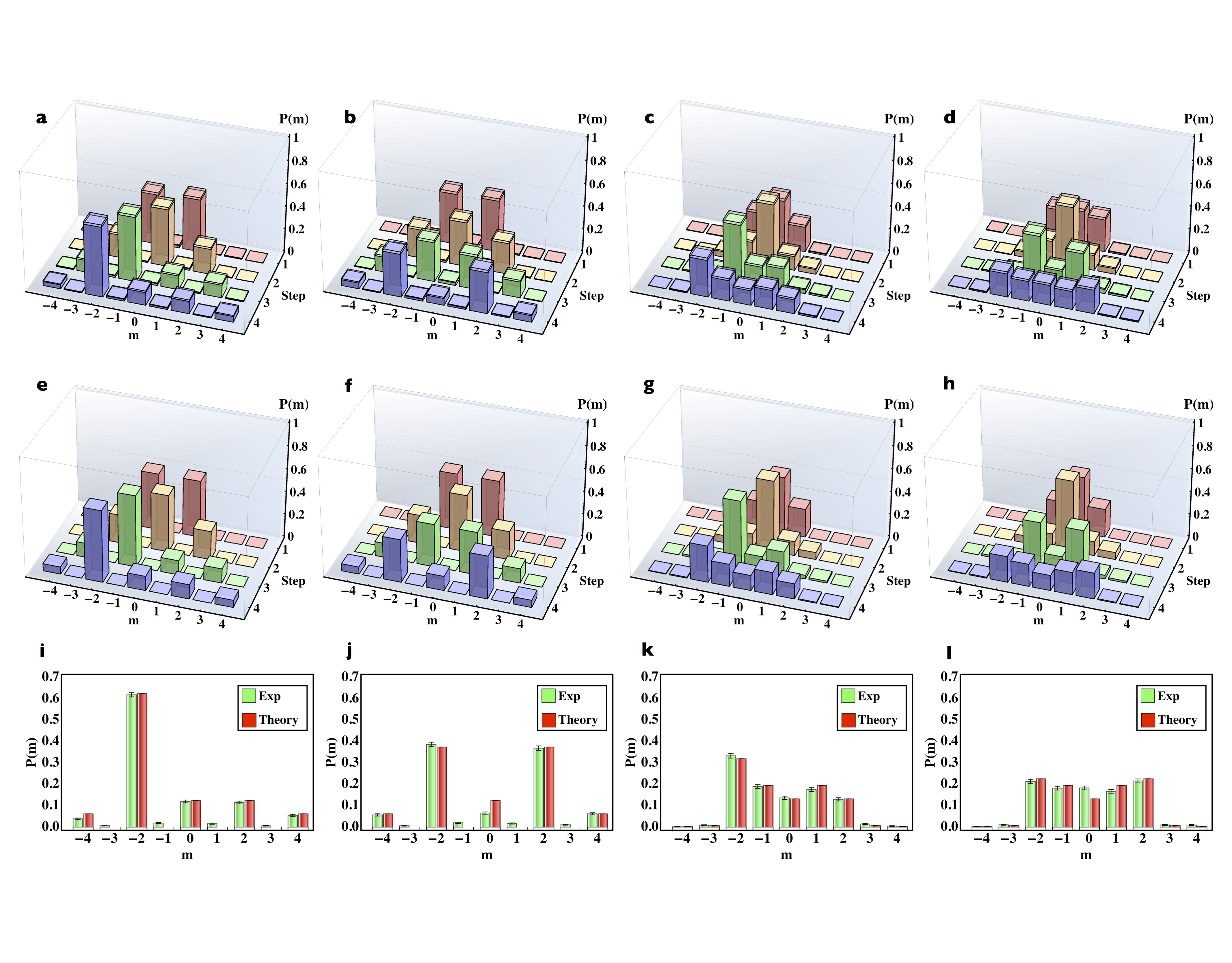}
\caption{Four-step quantum walk for a single photon. a-d) Experimental results, including both intermediate and final probabilities for different OAM states in the evolution (summed over different polarizations). The intermediate probabilities at step $n$ are obtained by switching off all QPs that follow that step, that is setting $\delta=0$. Panels a) and b) refer to the standard case with two different input states for the coin subsystem, $(\alpha,\beta)= (0,1)$ and $1/\sqrt{2} (1,i)$, respectively. c) and d) refer to the hybrid case with $\delta=\pi/2$, with the same initial coin-states. e-h) Corresponding theoretical predictions. i-l) Comparison of measured and predicted final probabilities. Poissonian statistical uncertainties at plus-or-minus one standard deviation are shown by error bars in panels i-n and as transparent-volumes in panels a-e. The similarities between experimental and predicted OAM distributions are $(94.7\pm0.4)\%$, $(93.4\pm0.5)\%$, $(99.7\pm0.1)\%$ and $(99.2\pm0.2)\%$, respectively. Panels on the same column refer to the same configuration and initial states.}
\label{fig:data1}
\end{figure*}
In our experiment, the step operator $\hat U$ is hence implemented by a sequence of a QWP, a QP, and a HWP. The QPs have $q=1/2$, so as to induce OAM shifts of $\pm1$. Due to reflection losses (mainly at the QP, which is not antireflection-coated), each step has a transmission efficiency of 86\% (but adding an antireflection coating will improve this value to $>95$\%). The $n$-steps walk is then implemented by simply cascading a sequence of QWP-QP-HWP on the single optical axis of the system. In the implemented setup, the linear distance $d$ between adjacent steps is small compared to the Rayleigh range $z_R$ of the photons, i.e. $d/z_R\ll1$ (near-field regime), so as to avoid optical effects that would alter the nature of the simulated process; a detailed discussion is provided in the SI. The layout of the apparatus is shown in Fig.\ \ref{fig:layout}. At the input of the QW apparatus, a pair of indistinguishable photons is generated in the product state $\ket{H}\ket{V}$, where $H$ and $V$ stand for horizontal and vertical linear polarization. The photon pairs are generated by spontaneous parametric down-conversion in a $\beta$-barium borate nonlinear crystal cut for degenerate collinear type-II phase matching, pumped by frequency-duplicated laser pulses at a wavelength of $400$ nm at 140 mW of average power (the generation setup is not shown in  Fig.\ \ref{fig:layout}). Both photons of each pair are then coupled into the same single-mode optical fiber, thus setting $m=0$ and ensuring a high degree of spatial indistinguishability. At the exit of the fiber, the initial polarization of the two photons is recovered using a QWP-HWP set. Let us now consider first the single-photon experiments, while further below we will discuss the two-photon case.
%
%Data 2 photons%
%
\begin{figure*}[t]
\centering
\includegraphics[width=16cm]{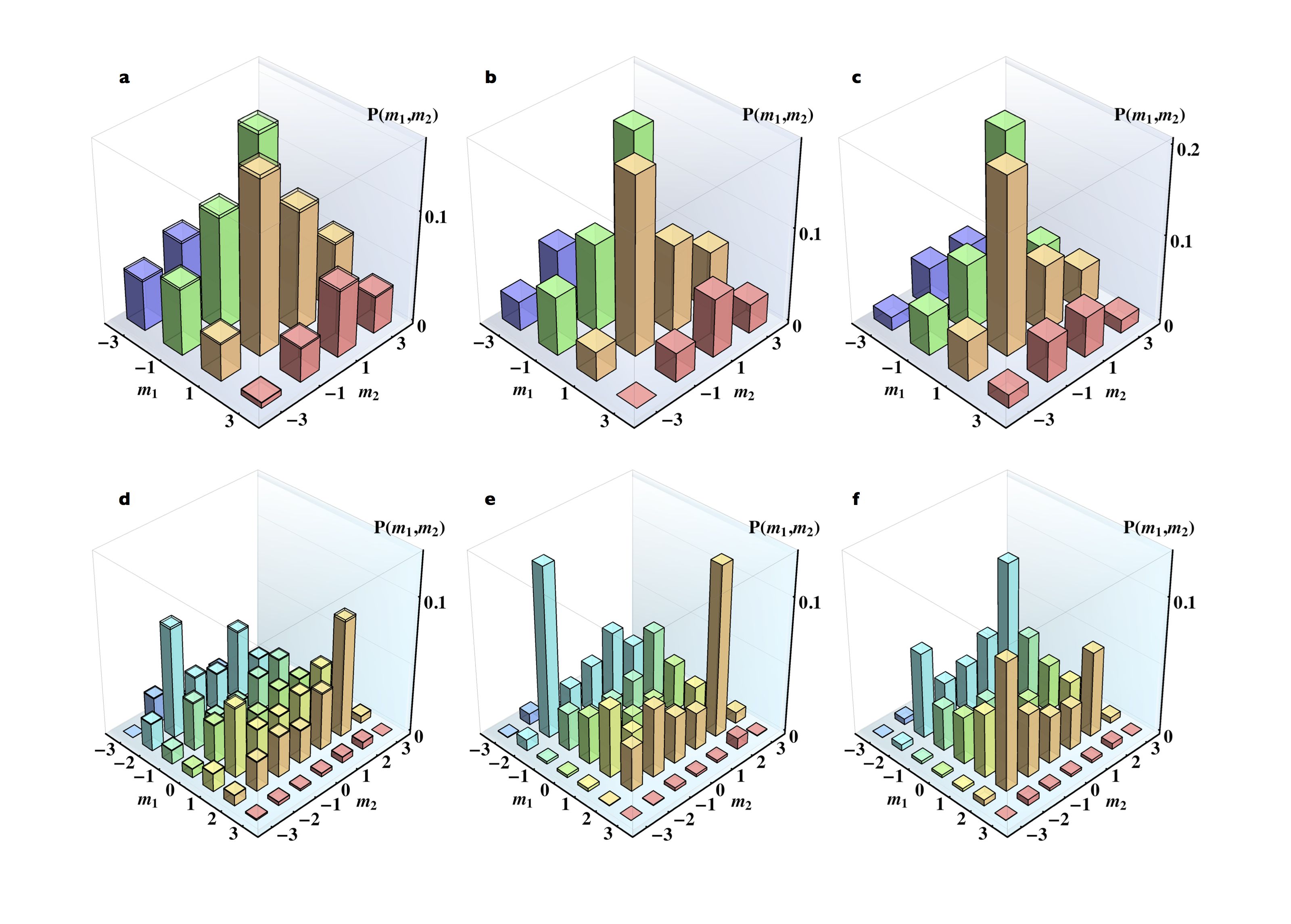}
\caption{Three-step quantum walk for two identical photons. In this case, only final OAM probabilities are shown (summed over different polarizations). a-c) Case of standard walk. a) Experimental results. Vertical bars represent estimated joint probabilities for the OAM of the two photons. Since the two measured photons detected after the BS splitting are physically equivalent, their counts are averaged together, so that $(m_1, m_2)$ and $(m_2, m_1)$ pairs actually refer to the same piece of data. Even values of $m_1$ and $m_2$ are not included, since they correspond to sites that cannot be occupied after an odd number of steps. b) Theoretical predictions for the case of indistinguishable photons. c) Theoretical predictions for the case of distinguishable photons, shown to highlight the effect of two-photon interferences (Hong-Ou-Mandel effect) in the final probabilities. It can be seen that the experimental results agree better with the theory for indistinguishable photons. This is further confirmed by the violation of corresponding inequalities, as discussed in the SI. d-f) Case of hybrid walk (with $\delta=\pi/2$). d), e) and f) refer respectively to experimental data, indistinguishable photon theory and distinguishable photon theory, as in the previous case. The specific hybrid walk implemented here is obtained without the HWP in each QW step, but this does not alter the nature of the process. In this case, the role of two-photon interferences appears to be stronger, so that the difference between e) and f) is more evident. Again, our experiment is in good agreement with the theory based on indistinguishable photons (see also the SI), proving that two-photon interferences are successfully implemented in our experiment. The similarities between experimental and predicted identical-photon distributions are $(98.2\pm0.1)\%$ and $(93.0\pm0.1)\%$ for the standard and the hybrid walk, respectively.}
\label{fig:data2}
\end{figure*}

To carry out a single-particle QW simulation, we split the two input photons with a polarizing beam splitter (PBS) and let the $H$-polarized photon only enter the QW setup. The $V$-polarized photon, reflected at the PBS, is sent directly to a detector and provides a trigger, so as to operate the QW simulation in a heralded single-photon quantum regime. The photon entering the QW setup is initially prepared in the arbitrarily polarized state $\ket{\psi_0}=\alpha\ket{L}+\beta\ket{R}$, where the two complex coefficients $\alpha$ and $\beta$ (such that $|\alpha|^2+|\beta|^2=1$) can be selected at will by a QWP-HWP set (apart from an unimportant global phase). The photon then undergoes the QW evolution and, at the exit, is analyzed in both polarization and OAM so as to determine the output probabilities. Details on these projective measurements in OAM are given in SI. In Fig.\ \ref{fig:data1} we report the experimental and predicted results relative to a 4-steps QW of a single photon, for two possible input polarization states, and both in the standard and hybrid configurations (two additional input polarization cases are given in Fig.\ S1 in the SI).

To investigate the simultaneous QW of two identical photons, both input photons were sent in the QW setup, after adjusting the input polarization to $\ket{R}\ket{L}$ (this is not the only possible choice, but it represents a typical case). At the exit of the QW cascade, we split the two photons with a beam splitter and analyze them both in polarization and OAM, so as to obtain the joint probability distribution (see Fig.\ \ref{fig:layout}). In Fig.\ \ref{fig:data2}, the results relative to a 3-steps QW are reported and compared with the theoretical predictions obtained for indistinguishable photons (taking into account also the effect of the beam splitter). The two distributions show a good qualitative agreement. The predictions for the case of distinguishable photons is also shown for comparison, to highlight the role of two-particle interferences in the final distributions. The measured distributions are also found to violate the characteristic inequalities that constrain the correlation distributions obtained with two classical light sources instead of two photons \cite{Peru10_Sci}, or with two distinguishable photons, as illustrated in the SI (Figs.\ S2 and S3). This proves that the measured correlations must be quantum and that they include the effect of multiparticle interferences.

To evaluate more quantitatively the agreement between measured and predicted probability distributions, $P(m)$ and $P'(m)$, we computed their ``similarity''\\
$S=\left(\sum_{m}\sqrt{P(m)P'(m)}\right)^2 / \left(\sum_{m}P(m)\sum_{m}P'(m)\right)$. In the case of the two-photon distributions, the index $m$ is replaced by the pair of OAM values $(m_1,m_2)$. As reported in the figure captions, the similarities were found to be always larger than $90\%$, thus confirming a good quantitative agreement between theory and experiment.

In conclusion, we have demonstrated a multi-photon quantum walk simulator based on single beam propagation through linear optical devices. The realized architecture is efficient and stable. Moreover, in contrast to other photonic QW implementations, the number of optical components employed scales only linearly with the number of steps, because at each step all OAM values are addressed simultaneously by a single optical element, and the element utilized transverse extension remains constant at each step. It must be noted however that this advantage in scaling remains valid only as long as the entire QW takes place in the optical near field, in which the beam cross-section size will remain approximately constant, while in the far field the transverse size of the optical components will have to increase with the OAM range. A current limitation of our approach is that the walk evolution cannot be position-dependent (that is, OAM-dependent), in contrast to other implementations\cite{Schr12_Sci,Cres13_NatPhot}. This limitation could be overcome in the future by introducing additional azimuthally-patterned optical elements and by exploiting also the radial beam coordinate, which couples with OAM in free propagation. On the other hand, our approach allows a very convenient and easy control of the evolution operator at each step, including the possibility of fully-automated fast switching of its properties. This may enable, for example, the simulation of a quantum system having a time-dependent Hamiltonian or that of a statistical ensemble of quantum systems with different Hamiltonians. Other potential future advantages of the present implementation include the relatively easy extension to the case in which the walker enters the system in an initial delocalized state\cite{Abal06_PRA}, case which has not been explored hitherto, and the possibility to carry out a full quantum tomography of the outgoing state, which is very challenging for standard interferometric implementations.

\newpage
%\noindent\textbf{REFERENCES}
%\bibliography{quantumrandomwalk}

\begin{thebibliography}{29}%
\makeatletter
\providecommand \@ifxundefined [1]{%
 \@ifx{#1\undefined}
}%
\providecommand \@ifnum [1]{%
 \ifnum #1\expandafter \@firstoftwo
 \else \expandafter \@secondoftwo
 \fi
}%
\providecommand \@ifx [1]{%
 \ifx #1\expandafter \@firstoftwo
 \else \expandafter \@secondoftwo
 \fi
}%
\providecommand \natexlab [1]{#1}%
\providecommand \enquote  [1]{``#1''}%
\providecommand \bibnamefont  [1]{#1}%
\providecommand \bibfnamefont [1]{#1}%
\providecommand \citenamefont [1]{#1}%
\providecommand \href@noop [0]{\@secondoftwo}%
\providecommand \href [0]{\begingroup \@sanitize@url \@href}%
\providecommand \@href[1]{\@@startlink{#1}\@@href}%
\providecommand \@@href[1]{\endgroup#1\@@endlink}%
\providecommand \@sanitize@url [0]{\catcode `\\12\catcode `\$12\catcode
  `\&12\catcode `\#12\catcode `\^12\catcode `\_12\catcode `\%12\relax}%
\providecommand \@@startlink[1]{}%
\providecommand \@@endlink[0]{}%
\providecommand \url  [0]{\begingroup\@sanitize@url \@url }%
\providecommand \@url [1]{\endgroup\@href {#1}{\urlprefix }}%
\providecommand \urlprefix  [0]{URL }%
\providecommand \Eprint [0]{\href }%
\providecommand \doibase [0]{http://dx.doi.org/}%
\providecommand \selectlanguage [0]{\@gobble}%
\providecommand \bibinfo  [0]{\@secondoftwo}%
\providecommand \bibfield  [0]{\@secondoftwo}%
\providecommand \translation [1]{[#1]}%
\providecommand \BibitemOpen [0]{}%
\providecommand \bibitemStop [0]{}%
\providecommand \bibitemNoStop [0]{.\EOS\space}%
\providecommand \EOS [0]{\spacefactor3000\relax}%
\providecommand \BibitemShut  [1]{\csname bibitem#1\endcsname}%
\let\auto@bib@innerbib\@empty
%</preamble>
\bibitem [{\citenamefont {Kempe}(2003)}]{Kemp03_ConPhys}%
  \BibitemOpen
  \bibfield  {author} {\bibinfo {author} {\bibfnamefont {J.}~\bibnamefont
  {Kempe}},\ }\bibfield  {title} {\enquote {\bibinfo {title} {Quantum random
  walks: An introductory overview},}\ }\href@noop {} {\bibfield  {journal}
  {\bibinfo  {journal} {Contemp. Phys.}\ }\textbf {\bibinfo {volume} {44}},\
  \bibinfo {pages} {307--327} (\bibinfo {year} {2003})}\BibitemShut {NoStop}%
\bibitem [{\citenamefont {Mohseni}\ \emph {et~al.}(2008)\citenamefont
  {Mohseni}, \citenamefont {Rebentrost}, \citenamefont {Lloyd},\ and\
  \citenamefont {Aspuru-Guzik}}]{Mohs08_JCP}%
  \BibitemOpen
  \bibfield  {author} {\bibinfo {author} {\bibfnamefont {M.}~\bibnamefont
  {Mohseni}}, \bibinfo {author} {\bibfnamefont {P.}~\bibnamefont {Rebentrost}},
  \bibinfo {author} {\bibfnamefont {S.}~\bibnamefont {Lloyd}}, \ and\ \bibinfo
  {author} {\bibfnamefont {A.}~\bibnamefont {Aspuru-Guzik}},\ }\bibfield
  {title} {\enquote {\bibinfo {title} {Environment-assisted quantum walks in
  photosynthetic energy transfer},}\ }\href
  {http://scitation.aip.org/content/aip/journal/jcp/129/17/10.1063/1.3002335}
  {\bibfield  {journal} {\bibinfo  {journal} {J. Chem. Phys}\ }\textbf
  {\bibinfo {volume} {129}},\ \bibinfo {eid} {174106} (\bibinfo {year}
  {2008})}\BibitemShut {NoStop}%
\bibitem [{\citenamefont {Kitagawa}\ \emph {et~al.}(2012)\citenamefont
  {Kitagawa}, \citenamefont {Broome}, \citenamefont {Fedrizzi}, \citenamefont
  {Rudner}, \citenamefont {Berg}, \citenamefont {Kassal}, \citenamefont
  {Aspuru-Guzik}, \citenamefont {Demler},\ and\ \citenamefont
  {White}}]{Kita12_NatCom}%
  \BibitemOpen
  \bibfield  {author} {\bibinfo {author} {\bibfnamefont {T.}~\bibnamefont
  {Kitagawa}}, \bibinfo {author} {\bibfnamefont {M.~A.}\ \bibnamefont
  {Broome}}, \bibinfo {author} {\bibfnamefont {A.}~\bibnamefont {Fedrizzi}},
  \bibinfo {author} {\bibfnamefont {M.~S.}\ \bibnamefont {Rudner}}, \bibinfo
  {author} {\bibfnamefont {E.}~\bibnamefont {Berg}}, \bibinfo {author}
  {\bibfnamefont {I.}~\bibnamefont {Kassal}}, \bibinfo {author} {\bibfnamefont
  {A.}~\bibnamefont {Aspuru-Guzik}}, \bibinfo {author} {\bibfnamefont
  {E.}~\bibnamefont {Demler}}, \ and\ \bibinfo {author} {\bibfnamefont {A.~G.}\
  \bibnamefont {White}},\ }\bibfield  {title} {\enquote {\bibinfo {title}
  {Observation of topologically protected bound states in photonic quantum
  walks},}\ }\href@noop {} {\bibfield  {journal} {\bibinfo  {journal} {Nat.
  Commun.}\ }\textbf {\bibinfo {volume} {3}} (\bibinfo {year}
  {2012})}\BibitemShut {NoStop}%
\bibitem [{\citenamefont {Crespi}\ \emph {et~al.}(2013)\citenamefont {Crespi},
  \citenamefont {Osellame}, \citenamefont {Ramponi}, \citenamefont
  {Giovannetti}, \citenamefont {Fazio}, \citenamefont {Sansoni}, \citenamefont
  {De~Nicola}, \citenamefont {Sciarrino},\ and\ \citenamefont
  {Mataloni}}]{Cres13_NatPhot}%
  \BibitemOpen
  \bibfield  {author} {\bibinfo {author} {\bibfnamefont {A.}~\bibnamefont
  {Crespi}}, \bibinfo {author} {\bibfnamefont {R.}~\bibnamefont {Osellame}},
  \bibinfo {author} {\bibfnamefont {R.}~\bibnamefont {Ramponi}}, \bibinfo
  {author} {\bibfnamefont {V.}~\bibnamefont {Giovannetti}}, \bibinfo {author}
  {\bibfnamefont {R.}~\bibnamefont {Fazio}}, \bibinfo {author} {\bibfnamefont
  {L.}~\bibnamefont {Sansoni}}, \bibinfo {author} {\bibfnamefont
  {F.}~\bibnamefont {De~Nicola}}, \bibinfo {author} {\bibfnamefont
  {F.}~\bibnamefont {Sciarrino}}, \ and\ \bibinfo {author} {\bibfnamefont
  {P.}~\bibnamefont {Mataloni}},\ }\bibfield  {title} {\enquote {\bibinfo
  {title} {Anderson localization of entangled photons in an integrated quantum
  walk},}\ }\href@noop {} {\bibfield  {journal} {\bibinfo  {journal} {Nat.
  Phot.}\ }\textbf {\bibinfo {volume} {7}},\ \bibinfo {pages} {322--328}
  (\bibinfo {year} {2013})}\BibitemShut {NoStop}%
\bibitem [{\citenamefont {Abal}\ \emph {et~al.}(2006)\citenamefont {Abal},
  \citenamefont {Siri}, \citenamefont {Romanelli},\ and\ \citenamefont
  {Donangelo}}]{Abal06_PRA}%
  \BibitemOpen
  \bibfield  {author} {\bibinfo {author} {\bibfnamefont {G.}~\bibnamefont
  {Abal}}, \bibinfo {author} {\bibfnamefont {R.}~\bibnamefont {Siri}}, \bibinfo
  {author} {\bibfnamefont {A.}~\bibnamefont {Romanelli}}, \ and\ \bibinfo
  {author} {\bibfnamefont {R.}~\bibnamefont {Donangelo}},\ }\bibfield  {title}
  {\enquote {\bibinfo {title} {Quantum walk on the line: Entanglement and
  nonlocal initial conditions},}\ }\href {\doibase 10.1103/PhysRevA.73.042302}
  {\bibfield  {journal} {\bibinfo  {journal} {Phys. Rev. A}\ }\textbf {\bibinfo
  {volume} {73}},\ \bibinfo {pages} {042302} (\bibinfo {year}
  {2006})}\BibitemShut {NoStop}%
\bibitem [{\citenamefont {Vieira}\ \emph {et~al.}(2013)\citenamefont {Vieira},
  \citenamefont {Amorim},\ and\ \citenamefont {Rigolin}}]{Viei13_PRL}%
  \BibitemOpen
  \bibfield  {author} {\bibinfo {author} {\bibfnamefont {R.}~\bibnamefont
  {Vieira}}, \bibinfo {author} {\bibfnamefont {E.~P.~M.}\ \bibnamefont
  {Amorim}}, \ and\ \bibinfo {author} {\bibfnamefont {G.}~\bibnamefont
  {Rigolin}},\ }\bibfield  {title} {\enquote {\bibinfo {title} {Dynamically
  disordered quantum walk as a maximal entanglement generator},}\ }\href
  {\doibase 10.1103/PhysRevLett.111.180503} {\bibfield  {journal} {\bibinfo
  {journal} {Phys. Rev. Lett.}\ }\textbf {\bibinfo {volume} {111}},\ \bibinfo
  {pages} {180503} (\bibinfo {year} {2013})}\BibitemShut {NoStop}%
\bibitem [{\citenamefont {A.M.Childs}(2009)}]{Chil09_PRL}%
  \BibitemOpen
  \bibfield  {author} {\bibinfo {author} {\bibnamefont {A.M.Childs}},\
  }\bibfield  {title} {\enquote {\bibinfo {title} {Universal computation by
  quantum walks},}\ }\href@noop {} {\bibfield  {journal} {\bibinfo  {journal}
  {Phys. Rev .Lett.}\ }\textbf {\bibinfo {volume} {102}},\ \bibinfo {pages}
  {180501} (\bibinfo {year} {2009})}\BibitemShut {NoStop}%
\bibitem [{\citenamefont {Lovett}\ \emph {et~al.}(2010)\citenamefont {Lovett},
  \citenamefont {Cooper}, \citenamefont {Everitt}, \citenamefont {Trevers},\
  and\ \citenamefont {Kendon}}]{Love10_PRA}%
  \BibitemOpen
  \bibfield  {author} {\bibinfo {author} {\bibfnamefont {N.~B.}\ \bibnamefont
  {Lovett}}, \bibinfo {author} {\bibfnamefont {S.}~\bibnamefont {Cooper}},
  \bibinfo {author} {\bibfnamefont {M.}~\bibnamefont {Everitt}}, \bibinfo
  {author} {\bibfnamefont {M.}~\bibnamefont {Trevers}}, \ and\ \bibinfo
  {author} {\bibfnamefont {V.}~\bibnamefont {Kendon}},\ }\bibfield  {title}
  {\enquote {\bibinfo {title} {Universal quantum computation using the
  discrete-time quantum walk},}\ }\href {\doibase 10.1103/PhysRevA.81.042330}
  {\bibfield  {journal} {\bibinfo  {journal} {Phys. Rev. A}\ }\textbf {\bibinfo
  {volume} {81}},\ \bibinfo {pages} {042330} (\bibinfo {year}
  {2010})}\BibitemShut {NoStop}%
\bibitem [{\citenamefont {Aspuru-Guzik}\ and\ \citenamefont
  {Walther}(2012)}]{Walt12_NatPhot}%
  \BibitemOpen
  \bibfield  {author} {\bibinfo {author} {\bibfnamefont {A.}~\bibnamefont
  {Aspuru-Guzik}}\ and\ \bibinfo {author} {\bibfnamefont {P.}~\bibnamefont
  {Walther}},\ }\bibfield  {title} {\enquote {\bibinfo {title} {Photonic
  quantum simulators},}\ }\href@noop {} {\bibfield  {journal} {\bibinfo
  {journal} {Nat. Phys.}\ }\textbf {\bibinfo {volume} {8}},\ \bibinfo {pages}
  {285--291} (\bibinfo {year} {2012})}\BibitemShut {NoStop}%
\bibitem [{\citenamefont {Feynman}(1982)}]{Feyn82}%
  \BibitemOpen
  \bibfield  {author} {\bibinfo {author} {\bibfnamefont {R.}~\bibnamefont
  {Feynman}},\ }\bibfield  {title} {\enquote {\bibinfo {title} {Simulating
  physics with computers},}\ }\href {\doibase 10.1007/BF02650179} {\bibfield
  {journal} {\bibinfo  {journal} {Int. J. Theor. Phys.}\ }\textbf {\bibinfo
  {volume} {21}},\ \bibinfo {pages} {467--488} (\bibinfo {year}
  {1982})}\BibitemShut {NoStop}%
\bibitem [{\citenamefont {Knight}\ \emph {et~al.}(2003)\citenamefont {Knight},
  \citenamefont {Rold\'an},\ and\ \citenamefont {Sipe}}]{Knig06_PRA}%
  \BibitemOpen
  \bibfield  {author} {\bibinfo {author} {\bibfnamefont {P.~L.}\ \bibnamefont
  {Knight}}, \bibinfo {author} {\bibfnamefont {E.}~\bibnamefont {Rold\'an}}, \
  and\ \bibinfo {author} {\bibfnamefont {J.~E.}\ \bibnamefont {Sipe}},\
  }\bibfield  {title} {\enquote {\bibinfo {title} {Quantum walk on the line as
  an interference phenomenon},}\ }\href {\doibase 10.1103/PhysRevA.68.020301}
  {\bibfield  {journal} {\bibinfo  {journal} {Phys. Rev. A}\ }\textbf {\bibinfo
  {volume} {68}},\ \bibinfo {pages} {020301} (\bibinfo {year}
  {2003})}\BibitemShut {NoStop}%
\bibitem [{\citenamefont {N.Shenvi}\ \emph {et~al.}(2003)\citenamefont
  {N.Shenvi}, \citenamefont {J.Kempe},\ and\ \citenamefont
  {B.Whaley}}]{Shen03_PRA}%
  \BibitemOpen
  \bibfield  {author} {\bibinfo {author} {\bibnamefont {N.Shenvi}}, \bibinfo
  {author} {\bibnamefont {J.Kempe}}, \ and\ \bibinfo {author} {\bibnamefont
  {B.Whaley}},\ }\bibfield  {title} {\enquote {\bibinfo {title} {Quantum
  random-walk search algorithm},}\ }\href@noop {} {\bibfield  {journal}
  {\bibinfo  {journal} {Phys. Rev. A}\ }\textbf {\bibinfo {volume} {67}},\
  \bibinfo {pages} {052307} (\bibinfo {year} {2003})}\BibitemShut {NoStop}%
\bibitem [{\citenamefont {Poto\ifmmode~\check{c}\else \v{c}\fi{}ek}\ \emph
  {et~al.}(2009)\citenamefont {Poto\ifmmode~\check{c}\else \v{c}\fi{}ek},
  \citenamefont {G\'abris}, \citenamefont {Kiss},\ and\ \citenamefont
  {Jex}}]{Poto09_PRA}%
  \BibitemOpen
  \bibfield  {author} {\bibinfo {author} {\bibfnamefont {V.}~\bibnamefont
  {Poto\ifmmode~\check{c}\else \v{c}\fi{}ek}}, \bibinfo {author} {\bibfnamefont
  {A.}~\bibnamefont {G\'abris}}, \bibinfo {author} {\bibfnamefont
  {T.}~\bibnamefont {Kiss}}, \ and\ \bibinfo {author} {\bibfnamefont
  {I.}~\bibnamefont {Jex}},\ }\bibfield  {title} {\enquote {\bibinfo {title}
  {Optimized quantum random-walk search algorithms on the hypercube},}\ }\href
  {\doibase 10.1103/PhysRevA.79.012325} {\bibfield  {journal} {\bibinfo
  {journal} {Phys. Rev. A}\ }\textbf {\bibinfo {volume} {79}},\ \bibinfo
  {pages} {012325} (\bibinfo {year} {2009})}\BibitemShut {NoStop}%
\bibitem [{\citenamefont {Schmitz}\ \emph {et~al.}(2009)\citenamefont
  {Schmitz}, \citenamefont {Matjeschk}, \citenamefont {Schneider},
  \citenamefont {Glueckert}, \citenamefont {Enderlein}, \citenamefont {Huber},\
  and\ \citenamefont {Schaetz}}]{Schm09_PRL}%
  \BibitemOpen
  \bibfield  {author} {\bibinfo {author} {\bibfnamefont {H.}~\bibnamefont
  {Schmitz}}, \bibinfo {author} {\bibfnamefont {R.}~\bibnamefont {Matjeschk}},
  \bibinfo {author} {\bibfnamefont {C.}~\bibnamefont {Schneider}}, \bibinfo
  {author} {\bibfnamefont {J.}~\bibnamefont {Glueckert}}, \bibinfo {author}
  {\bibfnamefont {M.}~\bibnamefont {Enderlein}}, \bibinfo {author}
  {\bibfnamefont {T.}~\bibnamefont {Huber}}, \ and\ \bibinfo {author}
  {\bibfnamefont {T.}~\bibnamefont {Schaetz}},\ }\bibfield  {title} {\enquote
  {\bibinfo {title} {Quantum walk of a trapped ion in phase space},}\ }\href
  {\doibase 10.1103/PhysRevLett.103.090504} {\bibfield  {journal} {\bibinfo
  {journal} {Phys. Rev. Lett.}\ }\textbf {\bibinfo {volume} {103}},\ \bibinfo
  {pages} {090504} (\bibinfo {year} {2009})}\BibitemShut {NoStop}%
\bibitem [{\citenamefont {Z\"ahringer}\ \emph {et~al.}(2010)\citenamefont
  {Z\"ahringer}, \citenamefont {Kirchmair}, \citenamefont {Gerritsma},
  \citenamefont {Solano}, \citenamefont {Blatt},\ and\ \citenamefont
  {Roos}}]{Zahr10_PRL}%
  \BibitemOpen
  \bibfield  {author} {\bibinfo {author} {\bibfnamefont {F.}~\bibnamefont
  {Z\"ahringer}}, \bibinfo {author} {\bibfnamefont {G.}~\bibnamefont
  {Kirchmair}}, \bibinfo {author} {\bibfnamefont {R.}~\bibnamefont
  {Gerritsma}}, \bibinfo {author} {\bibfnamefont {E.}~\bibnamefont {Solano}},
  \bibinfo {author} {\bibfnamefont {R.}~\bibnamefont {Blatt}}, \ and\ \bibinfo
  {author} {\bibfnamefont {C.~F.}\ \bibnamefont {Roos}},\ }\bibfield  {title}
  {\enquote {\bibinfo {title} {Realization of a quantum walk with one and two
  trapped ions},}\ }\href {\doibase 10.1103/PhysRevLett.104.100503} {\bibfield
  {journal} {\bibinfo  {journal} {Phys. Rev. Lett.}\ }\textbf {\bibinfo
  {volume} {104}},\ \bibinfo {pages} {100503} (\bibinfo {year}
  {2010})}\BibitemShut {NoStop}%
\bibitem [{\citenamefont {Karski}\ \emph {et~al.}(2009)\citenamefont {Karski},
  \citenamefont {F{\"o}rster}, \citenamefont {Choi}, \citenamefont {Steffen},
  \citenamefont {Alt}, \citenamefont {Meschede},\ and\ \citenamefont
  {Widera}}]{Kars09_Sci}%
  \BibitemOpen
  \bibfield  {author} {\bibinfo {author} {\bibfnamefont {M.}~\bibnamefont
  {Karski}}, \bibinfo {author} {\bibfnamefont {L.}~\bibnamefont {F{\"o}rster}},
  \bibinfo {author} {\bibfnamefont {J.M.}\ \bibnamefont {Choi}}, \bibinfo
  {author} {\bibfnamefont {A.}~\bibnamefont {Steffen}}, \bibinfo {author}
  {\bibfnamefont {W.}~\bibnamefont {Alt}}, \bibinfo {author} {\bibfnamefont
  {D.}~\bibnamefont {Meschede}}, \ and\ \bibinfo {author} {\bibfnamefont
  {A.}~\bibnamefont {Widera}},\ }\bibfield  {title} {\enquote {\bibinfo {title}
  {Quantum walk in position space with single optically trapped atoms},}\
  }\href {http://www.sciencemag.org/content/325/5937/174.abstract} {\bibfield
  {journal} {\bibinfo  {journal} {Science}\ }\textbf {\bibinfo {volume}
  {325}},\ \bibinfo {pages} {174--177} (\bibinfo {year} {2009})}\BibitemShut
  {NoStop}%
\bibitem [{\citenamefont {Ryan}\ \emph {et~al.}(2005)\citenamefont {Ryan},
  \citenamefont {Laforest}, \citenamefont {Boileau},\ and\ \citenamefont
  {Laflamme}}]{Ryan05_PRA}%
  \BibitemOpen
  \bibfield  {author} {\bibinfo {author} {\bibfnamefont {C.~A.}\ \bibnamefont
  {Ryan}}, \bibinfo {author} {\bibfnamefont {M.}~\bibnamefont {Laforest}},
  \bibinfo {author} {\bibfnamefont {J.~C.}\ \bibnamefont {Boileau}}, \ and\
  \bibinfo {author} {\bibfnamefont {R.}~\bibnamefont {Laflamme}},\ }\bibfield
  {title} {\enquote {\bibinfo {title} {Experimental implementation of a
  discrete-time quantum random walk on an nmr quantum-information processor},}\
  }\href {\doibase 10.1103/PhysRevA.72.062317} {\bibfield  {journal} {\bibinfo
  {journal} {Phys. Rev. A}\ }\textbf {\bibinfo {volume} {72}},\ \bibinfo
  {pages} {062317} (\bibinfo {year} {2005})}\BibitemShut {NoStop}%
\bibitem [{\citenamefont {Zhang}\ \emph {et~al.}(2007)\citenamefont {Zhang},
  \citenamefont {Ren}, \citenamefont {Zou}, \citenamefont {Liu}, \citenamefont
  {Huang},\ and\ \citenamefont {Guo}}]{Zhan07_PRA}%
  \BibitemOpen
  \bibfield  {author} {\bibinfo {author} {\bibfnamefont {P.}~\bibnamefont
  {Zhang}}, \bibinfo {author} {\bibfnamefont {X.F.}\ \bibnamefont {Ren}},
  \bibinfo {author} {\bibfnamefont {X.B.}\ \bibnamefont {Zou}}, \bibinfo
  {author} {\bibfnamefont {B.H.}\ \bibnamefont {Liu}}, \bibinfo {author}
  {\bibfnamefont {Y.F.}\ \bibnamefont {Huang}}, \ and\ \bibinfo {author}
  {\bibfnamefont {G.C.}\ \bibnamefont {Guo}},\ }\bibfield  {title} {\enquote
  {\bibinfo {title} {Demonstration of one-dimensional quantum random walks
  using orbital angular momentum of photons},}\ }\href@noop {} {\bibfield
  {journal} {\bibinfo  {journal} {Phys. Rev. A}\ }\textbf {\bibinfo {volume}
  {75}},\ \bibinfo {pages} {052310} (\bibinfo {year} {2007})}\BibitemShut
  {NoStop}%
\bibitem [{\citenamefont {Broome}\ \emph {et~al.}(2010)\citenamefont {Broome},
  \citenamefont {Fedrizzi}, \citenamefont {Lanyon}, \citenamefont {Kassal},
  \citenamefont {Aspuru-Guzik},\ and\ \citenamefont {White}}]{Broo10_PRL}%
  \BibitemOpen
  \bibfield  {author} {\bibinfo {author} {\bibfnamefont {M.~A.}\ \bibnamefont
  {Broome}}, \bibinfo {author} {\bibfnamefont {A.}~\bibnamefont {Fedrizzi}},
  \bibinfo {author} {\bibfnamefont {B.~P.}\ \bibnamefont {Lanyon}}, \bibinfo
  {author} {\bibfnamefont {I.}~\bibnamefont {Kassal}}, \bibinfo {author}
  {\bibfnamefont {A.}~\bibnamefont {Aspuru-Guzik}}, \ and\ \bibinfo {author}
  {\bibfnamefont {A.~G.}\ \bibnamefont {White}},\ }\bibfield  {title} {\enquote
  {\bibinfo {title} {Discrete single-photon quantum walks with tunable
  decoherence},}\ }\href@noop {} {\bibfield  {journal} {\bibinfo  {journal}
  {Phys. Rev .Lett.}\ }\textbf {\bibinfo {volume} {104}},\ \bibinfo {pages}
  {153602} (\bibinfo {year} {2010})}\BibitemShut {NoStop}%
\bibitem [{\citenamefont {Schreiber}\ \emph {et~al.}(2010)\citenamefont
  {Schreiber}, \citenamefont {Cassemiro}, \citenamefont {Potocek},
  \citenamefont {G\'abris}, \citenamefont {Mosley}, \citenamefont {Andersson},
  \citenamefont {Jex},\ and\ \citenamefont {Silberhorn}}]{Schr10_PRL}%
  \BibitemOpen
  \bibfield  {author} {\bibinfo {author} {\bibfnamefont {A.}~\bibnamefont
  {Schreiber}}, \bibinfo {author} {\bibfnamefont {K.N.}\ \bibnamefont
  {Cassemiro}}, \bibinfo {author} {\bibfnamefont {V.}~\bibnamefont {Potocek}},
  \bibinfo {author} {\bibfnamefont {A.}~\bibnamefont {G\'abris}}, \bibinfo
  {author} {\bibfnamefont {P.~J.}\ \bibnamefont {Mosley}}, \bibinfo {author}
  {\bibfnamefont {E.}~\bibnamefont {Andersson}}, \bibinfo {author}
  {\bibfnamefont {I.}~\bibnamefont {Jex}}, \ and\ \bibinfo {author}
  {\bibfnamefont {Ch.}\ \bibnamefont {Silberhorn}},\ }\bibfield  {title}
  {\enquote {\bibinfo {title} {Photons walking the line: A quantum walk with
  adjustable coin operations},}\ }\href@noop {} {\bibfield  {journal} {\bibinfo
   {journal} {Phys. Rev .Lett.}\ }\textbf {\bibinfo {volume} {104}},\ \bibinfo
  {pages} {050502} (\bibinfo {year} {2010})}\BibitemShut {NoStop}%
\bibitem [{\citenamefont {A.Peruzzo}\ \emph {et~al.}(2010)\citenamefont
  {A.Peruzzo}, \citenamefont {M.Lobino}, \citenamefont {J.C.F.Matthews},
  \citenamefont {N.Matsuda}, \citenamefont {A.Politi}, \citenamefont
  {K.Poulios}, \citenamefont {X.Zhou}, \citenamefont {Y.Lahini}, \citenamefont
  {N.Ismail}, \citenamefont {Worhoff}, \citenamefont {Y.Bromberg},
  \citenamefont {Y.Silberberg}, \citenamefont {M.G.Thompson},\ and\
  \citenamefont {O'Brien}}]{Peru10_Sci}%
  \BibitemOpen
  \bibfield  {author} {\bibinfo {author} {\bibnamefont {A.Peruzzo}}, \bibinfo
  {author} {\bibnamefont {M.Lobino}}, \bibinfo {author} {\bibnamefont
  {J.C.F.Matthews}}, \bibinfo {author} {\bibnamefont {N.Matsuda}}, \bibinfo
  {author} {\bibnamefont {A.Politi}}, \bibinfo {author} {\bibnamefont
  {K.Poulios}}, \bibinfo {author} {\bibnamefont {X.Zhou}}, \bibinfo {author}
  {\bibnamefont {Y.Lahini}}, \bibinfo {author} {\bibnamefont {N.Ismail}},
  \bibinfo {author} {\bibfnamefont {K.}~\bibnamefont {Worhoff}}, \bibinfo
  {author} {\bibnamefont {Y.Bromberg}}, \bibinfo {author} {\bibnamefont
  {Y.Silberberg}}, \bibinfo {author} {\bibnamefont {M.G.Thompson}}, \ and\
  \bibinfo {author} {\bibfnamefont {J.L.}\ \bibnamefont {O'Brien}},\ }\bibfield
   {title} {\enquote {\bibinfo {title} {Quantum walks of correlated photons},}\
  }\href@noop {} {\bibfield  {journal} {\bibinfo  {journal} {Science}\ }\textbf
  {\bibinfo {volume} {329}},\ \bibinfo {pages} {1500--1503} (\bibinfo {year}
  {2010})}\BibitemShut {NoStop}%
\bibitem [{\citenamefont {Sansoni}\ \emph {et~al.}(2012)\citenamefont
  {Sansoni}, \citenamefont {Sciarrino}, \citenamefont {Vallone}, \citenamefont
  {Mataloni}, \citenamefont {Crespi}, \citenamefont {Ramponi},\ and\
  \citenamefont {Osellame}}]{Sans12_PRL}%
  \BibitemOpen
  \bibfield  {author} {\bibinfo {author} {\bibfnamefont {L.}~\bibnamefont
  {Sansoni}}, \bibinfo {author} {\bibfnamefont {F.}~\bibnamefont {Sciarrino}},
  \bibinfo {author} {\bibfnamefont {G.}~\bibnamefont {Vallone}}, \bibinfo
  {author} {\bibfnamefont {P.}~\bibnamefont {Mataloni}}, \bibinfo {author}
  {\bibfnamefont {A.}~\bibnamefont {Crespi}}, \bibinfo {author} {\bibfnamefont
  {R.}~\bibnamefont {Ramponi}}, \ and\ \bibinfo {author} {\bibfnamefont
  {R.}~\bibnamefont {Osellame}},\ }\bibfield  {title} {\enquote {\bibinfo
  {title} {Two-particle bosonic-fermionic quantum walk via integrated
  photonics},}\ }\href@noop {} {\bibfield  {journal} {\bibinfo  {journal}
  {Phys. Rev. Lett.}\ }\textbf {\bibinfo {volume} {108}},\ \bibinfo {pages}
  {010502} (\bibinfo {year} {2012})}\BibitemShut {NoStop}%
\bibitem [{\citenamefont {Owens}\ \emph {et~al.}(2011)\citenamefont {Owens},
  \citenamefont {Broome}, \citenamefont {Biggerstaff}, \citenamefont {Goggin},
  \citenamefont {Fedrizzi}, \citenamefont {Linjordet}, \citenamefont {Ams},
  \citenamefont {Marshall}, \citenamefont {Twamley}, \citenamefont {Withford},\
  and\ \citenamefont {White}}]{Owen11_NJP}%
  \BibitemOpen
  \bibfield  {author} {\bibinfo {author} {\bibfnamefont {J.~O.}\ \bibnamefont
  {Owens}}, \bibinfo {author} {\bibfnamefont {M.~A.}\ \bibnamefont {Broome}},
  \bibinfo {author} {\bibfnamefont {D.~N.}\ \bibnamefont {Biggerstaff}},
  \bibinfo {author} {\bibfnamefont {M.~E.}\ \bibnamefont {Goggin}}, \bibinfo
  {author} {\bibfnamefont {A.}~\bibnamefont {Fedrizzi}}, \bibinfo {author}
  {\bibfnamefont {T.}~\bibnamefont {Linjordet}}, \bibinfo {author}
  {\bibfnamefont {M.}~\bibnamefont {Ams}}, \bibinfo {author} {\bibfnamefont
  {G.~D.}\ \bibnamefont {Marshall}}, \bibinfo {author} {\bibfnamefont
  {J.}~\bibnamefont {Twamley}}, \bibinfo {author} {\bibfnamefont {M.~J.}\
  \bibnamefont {Withford}}, \ and\ \bibinfo {author} {\bibfnamefont {A.~G.}\
  \bibnamefont {White}},\ }\bibfield  {title} {\enquote {\bibinfo {title}
  {Two-photon quantum walks in an elliptical direct-write waveguide array},}\
  }\href@noop {} {\bibfield  {journal} {\bibinfo  {journal} {New J. Phys.}\
  }\textbf {\bibinfo {volume} {13}},\ \bibinfo {pages} {075003} (\bibinfo
  {year} {2011})}\BibitemShut {NoStop}%
\bibitem [{\citenamefont {Schreiber}\ \emph {et~al.}(2012)\citenamefont
  {Schreiber}, \citenamefont {G{\'a}bris}, \citenamefont {Rohde}, \citenamefont
  {Laiho}, \citenamefont {{\v S}tefa{\v n}{\'a}k}, \citenamefont {Poto{\v
  c}ek}, \citenamefont {Hamilton}, \citenamefont {Jex},\ and\ \citenamefont
  {Silberhorn}}]{Schr12_Sci}%
  \BibitemOpen
  \bibfield  {author} {\bibinfo {author} {\bibfnamefont {A.}~\bibnamefont
  {Schreiber}}, \bibinfo {author} {\bibfnamefont {A.}~\bibnamefont
  {G{\'a}bris}}, \bibinfo {author} {\bibfnamefont {P.~P.}\ \bibnamefont
  {Rohde}}, \bibinfo {author} {\bibfnamefont {K.}~\bibnamefont {Laiho}},
  \bibinfo {author} {\bibfnamefont {M.}~\bibnamefont {{\v S}tefa{\v n}{\'a}k}},
  \bibinfo {author} {\bibfnamefont {V.}~\bibnamefont {Poto{\v c}ek}}, \bibinfo
  {author} {\bibfnamefont {C.}~\bibnamefont {Hamilton}}, \bibinfo {author}
  {\bibfnamefont {I.}~\bibnamefont {Jex}}, \ and\ \bibinfo {author}
  {\bibfnamefont {C.}~\bibnamefont {Silberhorn}},\ }\bibfield  {title}
  {\enquote {\bibinfo {title} {A 2d quantum walk simulation of two-particle
  dynamics},}\ }\href {http://www.sciencemag.org/content/336/6077/55.abstract}
  {\bibfield  {journal} {\bibinfo  {journal} {Science}\ }\textbf {\bibinfo
  {volume} {336}},\ \bibinfo {pages} {55--58} (\bibinfo {year}
  {2012})}\BibitemShut {NoStop}%
\bibitem [{\citenamefont {Zhang}\ \emph {et~al.}(2010)\citenamefont {Zhang},
  \citenamefont {Liu}, \citenamefont {Liu}, \citenamefont {Li}, \citenamefont
  {Li},\ and\ \citenamefont {Guo}}]{Zhan10_PRA}%
  \BibitemOpen
  \bibfield  {author} {\bibinfo {author} {\bibfnamefont {P.}~\bibnamefont
  {Zhang}}, \bibinfo {author} {\bibfnamefont {B.H.}\ \bibnamefont {Liu}},
  \bibinfo {author} {\bibfnamefont {R.F}\ \bibnamefont {Liu}}, \bibinfo
  {author} {\bibfnamefont {H.R.}\ \bibnamefont {Li}}, \bibinfo {author}
  {\bibfnamefont {F.L.}\ \bibnamefont {Li}}, \ and\ \bibinfo {author}
  {\bibfnamefont {G.C.}\ \bibnamefont {Guo}},\ }\bibfield  {title} {\enquote
  {\bibinfo {title} {Implementation of one-dimensional quantum walks on
  spin-orbital angular momentum space of photons},}\ }\href@noop {} {\bibfield
  {journal} {\bibinfo  {journal} {Phys. Rev. A}\ }\textbf {\bibinfo {volume}
  {81}},\ \bibinfo {pages} {052322} (\bibinfo {year} {2010})}\BibitemShut
  {NoStop}%
\bibitem [{\citenamefont {Goyal}\ \emph {et~al.}(2013)\citenamefont {Goyal},
  \citenamefont {Roux}, \citenamefont {Forbes},\ and\ \citenamefont
  {Konrad}}]{Goya13_PRL}%
  \BibitemOpen
  \bibfield  {author} {\bibinfo {author} {\bibfnamefont {S.~K.}\ \bibnamefont
  {Goyal}}, \bibinfo {author} {\bibfnamefont {F.~S.}\ \bibnamefont {Roux}},
  \bibinfo {author} {\bibfnamefont {A.}~\bibnamefont {Forbes}}, \ and\ \bibinfo
  {author} {\bibfnamefont {T.}~\bibnamefont {Konrad}},\ }\bibfield  {title}
  {\enquote {\bibinfo {title} {Implementing quantum walks using orbital angular
  momentum of classical light},}\ }\href {\doibase
  10.1103/PhysRevLett.110.263602} {\bibfield  {journal} {\bibinfo  {journal}
  {Phys. Rev. Lett.}\ }\textbf {\bibinfo {volume} {110}},\ \bibinfo {pages}
  {263602} (\bibinfo {year} {2013})}\BibitemShut {NoStop}%
\bibitem [{\citenamefont {L.Marrucci}\ \emph {et~al.}(2006)\citenamefont
  {L.Marrucci}, \citenamefont {C.Manzo},\ and\ \citenamefont
  {D.Paparo}}]{Marr06_PRL}%
  \BibitemOpen
  \bibfield  {author} {\bibinfo {author} {\bibnamefont {L.Marrucci}}, \bibinfo
  {author} {\bibnamefont {C.Manzo}}, \ and\ \bibinfo {author} {\bibnamefont
  {D.Paparo}},\ }\bibfield  {title} {\enquote {\bibinfo {title} {Optical
  spin-to-orbital angular momentum conversion in inhomogeneous anisotropic
  media},}\ }\href@noop {} {\bibfield  {journal} {\bibinfo  {journal} {Phys.
  Rev .Lett.}\ }\textbf {\bibinfo {volume} {97}},\ \bibinfo {pages} {163905}
  (\bibinfo {year} {2006})}\BibitemShut {NoStop}%
\bibitem [{\citenamefont {Marrucci}\ \emph {et~al.}(2011)\citenamefont
  {Marrucci}, \citenamefont {Karimi}, \citenamefont {Slussarenko},
  \citenamefont {Piccirillo}, \citenamefont {Santamato}, \citenamefont
  {Nagali},\ and\ \citenamefont {Sciarrino}}]{Marr11_JOpt}%
  \BibitemOpen
  \bibfield  {author} {\bibinfo {author} {\bibfnamefont {L.}~\bibnamefont
  {Marrucci}}, \bibinfo {author} {\bibfnamefont {E.}~\bibnamefont {Karimi}},
  \bibinfo {author} {\bibfnamefont {S.}~\bibnamefont {Slussarenko}}, \bibinfo
  {author} {\bibfnamefont {B.}~\bibnamefont {Piccirillo}}, \bibinfo {author}
  {\bibfnamefont {E.}~\bibnamefont {Santamato}}, \bibinfo {author}
  {\bibfnamefont {E.}~\bibnamefont {Nagali}}, \ and\ \bibinfo {author}
  {\bibfnamefont {F.}~\bibnamefont {Sciarrino}},\ }\bibfield  {title} {\enquote
  {\bibinfo {title} {Spin-to-orbital conversion of the angular momentum of
  light and its classical and quantum applications},}\ }\href@noop {}
  {\bibfield  {journal} {\bibinfo  {journal} {Journal of Optics}\ }\textbf
  {\bibinfo {volume} {13}},\ \bibinfo {pages} {064001} (\bibinfo {year}
  {2011})}\BibitemShut {NoStop}%
\bibitem [{\citenamefont {Piccirillo}\ \emph {et~al.}(2010)\citenamefont
  {Piccirillo}, \citenamefont {D'Ambrosio}, \citenamefont {Slussarenko},
  \citenamefont {Marrucci},\ and\ \citenamefont {Santamato}}]{Picc10_APL}%
  \BibitemOpen
  \bibfield  {author} {\bibinfo {author} {\bibfnamefont {B.}~\bibnamefont
  {Piccirillo}}, \bibinfo {author} {\bibfnamefont {V.}~\bibnamefont
  {D'Ambrosio}}, \bibinfo {author} {\bibfnamefont {S.}~\bibnamefont
  {Slussarenko}}, \bibinfo {author} {\bibfnamefont {L.}~\bibnamefont
  {Marrucci}}, \ and\ \bibinfo {author} {\bibfnamefont {E.}~\bibnamefont
  {Santamato}},\ }\bibfield  {title} {\enquote {\bibinfo {title} {Photon
  spin-to-orbital angular momentum conversion via an electrically tunable
  q-plate},}\ }\href@noop {} {\bibfield  {journal} {\bibinfo  {journal} {Appl.
  Phys. Lett.}\ }\textbf {\bibinfo {volume} {97}},\ \bibinfo {pages} {241104}
  (\bibinfo {year} {2010})}\BibitemShut {NoStop}%
\end{thebibliography}

\newpage

\begin{thebibliography}{99}

\bibitem{lorenzo:06s}
Marrucci, L., Manzo, C., \& Paparo, D. Optical spin-to-orbital angular momentum conversion in inhomogeneous anisotropic media, \textit{Phys.\ Rev.\ Lett.} {\bf 96}, 163905 (2006).

\bibitem{siegmans}
Siegman, A. E. \textit{Lasers} (University Science Books, 1986).

\bibitem{karimiol:09s}
Karimi, E., Piccirillo, B., Marrucci, L., \& Santamato, E. Light propagation in a birefringent plate with topological charge, \textit{Opt.\ Lett.} {\bf 34}, 1225--1227 (2009).

\bibitem{karimiol:07s}
Karimi, E., Zito, G., Piccirillo, B., Marrucci, L., \& Santamato, E. Hypergeometric-Gaussian modes, \textit{Opt.\ Lett.} {\bf 32}, 3053--3055 (2007).

\bibitem{mair:01s}
Mair, A., Vaziri, A., Weihs, G. \& Zeilinger, A. Entanglement of the orbital angular momentum states of photons, \textit{Nature} {\bf 412}, 313 (2001).

\bibitem{hammam:14s}
Qassim, H., Miatto, F. M., Torres, J. P., Padgett, M. J., Karimi, E., \& Boyd, R. W. Limitations to the determination of a Laguerre-Gauss spectrum via projective, phase-flattening measurement, \textit{arXiv}:1401.3512 (2014).

\bibitem{bromberg09s}
Bromberg, Y., Lahini, Y., Morandotti, R., \& Silberberg, Y. Quantum and Classical Correlations in Waveguide Lattices, \textit{Phys.\ Rev.\ Lett.} {\bf 102}, 253904 (2009).

\bibitem{Peru10s}
Peruzzo, A. \textit{et al.} Quantum Walks of Correlated Photons, \textit{Science} \textbf{329}, 1500-1503 (2010).
\end{thebibliography}
%
\vspace{1 EM}

%\newpage
\noindent\textbf{Supplementary Information} accompanies this manuscript.
\vspace{1 EM}

\noindent\textbf{Acknowledgments}

\noindent We thank Antonio Ramaglia and Marco Cilmo for lending some equipment and Pei Zhang for an early suggestion of the possibility to carry out a photonic quantum walk in OAM following the scheme proposed in his paper. This work was partly supported by the Future Emerging Technologies FET-Open Program, within the 7$^{th}$ Framework Programme of the European Commission, under Grant No.\ 255914, PHORBITECH. F.S. acknowledges also ERC Starting Grant 3D-QUEST (grant agreement no. 307783). E.K. acknowledges the support of the Canada Excellence Research Chairs (CERC) Program.
\vspace{1 EM}

\noindent\textbf{Author Contributions}

\noindent F.C., F.M, E.K., F.S., E.S. and L.M. devised the project and designed the experimental methodology. F.C. and F.M., with contributions from D.P. and C.d.L., carried out the experiment and analyzed the data. S.S. prepared the q-plates. F.C., F.M. and L.M. wrote the manuscript, with contributions from E.K. All authors discussed the results and contributed to refining the manuscript.
\vspace{1 EM}

\noindent\textbf{Author Information}

\noindent The authors declare no competing financial interests. Correspondence and requests for materials should be addressed to L.M. (lorenzo.marrucci@unina.it).
\clearpage

\onecolumngrid
\appendix
\textbf{Supplementary information for Photonic quantum walk in a single beam with twisted light}
\vspace{1 EM}
\section{The q-plate device}
A q-plate\cite{lorenzo:06s} (QP) consists of a thin slab of uniaxial birefringent nematic liquid crystal sandwiched between containing glasses, whose optical axis in the slab plane is engineered in a inhomogeneous pattern, according to the relation
\begin{equation}\label{met:oa}
\alpha(\phi)=q\,\phi+\alpha_0,
\end{equation}
where $\alpha$ is the angle formed by the optical axis with the reference (horizontal) axis, $\phi$ is the azimuthal coordinate in the transverse plane of the device, $q$ is the topological charge of the plate and $\alpha_0$ is the axis direction at $\phi=0$. When light passes through a QP, the angle $\alpha_0$ is responsible for a relative phase emerging between the various OAM components in the output state. Indeed, when $\alpha_0\neq0$, the action of the QP is described by the following equations
\begin{align}
\widehat Q^{\alpha_0}_\delta\ket{L,m}&=\cos{(\delta/2)}\ket{L,m}-i\sin{(\delta/2)}e^{i\,2\alpha_0}\ket{R,m+2q},\cr 
\widehat Q^{\alpha_0}_\delta\ket{R,m}&=\cos{(\delta/2)}\ket{L,m}-i\sin{(\delta/2)}e^{-i\,2\alpha_0}\ket{L,m-2q}.
\end{align}
which reduces to Eq. 4 of the main text when $\alpha_0=0$. A vanishing relative phase between these two terms is required to implement properly the operator $\hat U$ describing the quantum walk process. To achieve this, all QPs in our setup were suitably oriented to match the condition $\alpha_0=0$.
\section{Role of the radial modes}
Our realization of the quantum walk (QW) relies on the encoding of the quantum walker state in the transverse modes of light, in particular those associated with the azimuthal degree of freedom. For simplicity, the radial structure of the mode is not considered explicitly in our scheme. However, a full treatment of the optical process needs to take also the radial effects into account. Indeed, all optical devices used to manipulate the azimuthal structure and hence the OAM of light, including the QP, introduce unavoidably also some alteration of the radial profile of the beam, particularly when the susequent free propagation is taken into account.

In this context, we choose Laguerre-Gauss (LG) modes as the basis, as they provide a set of orthonormal solutions to the paraxial wave equation. LG modes are indexed by an integer $m$ and a positive integer $p$, which determine the beam azimuthal and radial structures, respectively. Using cylindrical coordinates $r,\phi,z$, this modes are given by 
\begin{align}
	\label{eq:lg}
	\mbox{LG}_{p,m}(r,\phi,z)&=\sqrt{\frac{ 2^{|m|+1}p!}{\pi w(z)^2\,(p+|m|)!}}\,\left(\frac{r}{w(z)}\right)^{|m|} e^{-\frac{r^2}{w(z)^2}} L_{p}^{|\ell|}\left(\frac{2r^2}{w(z)^2}\right)\,e^{\left(\frac{ikr^2}{2R(z)}\right)}\,e^{im\phi}\,e^{-i(2p+|m|+1)\arctan{\left(\frac{z}{z_R}\right)}},
\end{align}
where $k$ is the wave number, $w(z)=w_0\,\sqrt{1+(z/z_R)^2}$, $R(z)=z\left[1+(z/z_R)^2\right]$ and $z_R=kw_0^2/2$ are the beam radius, wavefront curvature radius and Rayleigh range, respectively, $w_0$ being the radius at the beam waist~\cite{siegman}. $L_{p}^{|\ell|}(x)$ are the generalized Laguerre polynomials.

As already discussed, the QP raises or lowers the OAM content of the incoming beam, according to its polarization state. Due to presence of the singularity at the origin, the QP also alters the radial index of the incoming beam. The details of these calculations are reported in Ref~\cite{karimiol:09s}. Based on this analysis and assuming a low birefringence of the liquid crystals a \emph{tuned} QP $(\delta=\pi)$ transforms a circularly polarized, e.g. left-handed, input LG$_{0,m} (r,\phi,0)$ beam into
\begin{align}\label{eq:qplate}
\widehat Q_{\pi}\mbox{LG}_{0,m}(r,z)\ket{L,m}&=-i\,\text{HyGG}_{|m|-|m+1|,m+1}(r,z)\ket{R,m+1},
\end{align}
where HyGG$_{p,m} (r,z)$ stands for the amplitude of Hypergeometric-Gauss (HyGG) modes~\cite{karimiol:07s} and the azimuthal term $e^{im\phi}$ has been replaced by the ket $\ket{m}$.  Introducing dimensionless coordinates $\rho=r/w_0$ and $\zeta=z/z_R$ these modes are given by
\begin{eqnarray}\label{eq:HyGG}
   \hbox{HyGG}_{pm}(\rho,\zeta)&=&i^{|m|+1}\sqrt{\frac{2^{p+|m|+1}}{\pi\Gamma(p+|m|+1)}} \frac{\Gamma\left(1+|m|+\frac{p}{2}\right)}{\Gamma\left(|m|+1\right)}\,\\
    &{\ensuremath\times}&\zeta^{\frac{p}{2}}(\zeta+i)^{-(1+|m|+\frac{p}{2})} \rho^{|m|}\,e^{-\frac{i\rho^2}{(\zeta+i)}}
   {}_1\!F_1\left(-\frac{p}{2},1+|m|;\frac{\rho^2}{\zeta(\zeta+i)}\right)\nonumber
\end{eqnarray}
where $\Gamma(x)$ is the gamma function and $F_1(a,b;x)$ is a confluent hypergeometric function. In order to determine the radial mode alteration introduced by the QP, we can expand the output beam in the LG modes basis, i.e. $\text{HyGG}_{|m|-|m+1|,m+1}=\sum_p{c_p\text{LG}_{p,m+1}}$~\cite{karimiol:07s}. The expansion coefficients are given by
\begin{eqnarray}\label{eq:coeff}
c_p=\sqrt{\frac{1}{p!\,m!\,\left(p+|m+1|\right)!}}\,\frac{\left(|m+1|+|m|\right)!\,\Gamma\left(p+\frac{|m+1|-|m|}{2}\right)}{\Gamma\left(\frac{|m+1|-|m|}{2}\right)}
\end{eqnarray}
\begin{table}
  \centering
  \caption{\label{tab:tab1} Power coefficients of the various $p$-index terms appearing in the expansion of the beam emerging from a QP (with $q=1/2$) in the LG modes basis, assuming at the input there is an $L$-polarized LG mode with $p=0$ and the given OAM $m$ value.}
 \begin{tabular*}{0.4\textwidth}{@{\extracolsep{\fill} }  c|cccc}
    \hline\hline
    OAM          &     $|c_0|^2$   & $|c_1|^2$  &     $|c_2|^2$   & $|c_3|^2$ \\ \hline\hline
    $m=0$       &   0.785  &  0.098  &  0.036   &  0.019 \\ \hline
    $m=1$       &   0.883  &  0.073  &  0.020   &  0.008 \\ \hline
    $m=2$       &   0.920  &  0.057  &  0.012   &  0.004 \\ \hline
    $m=3$       &   0.939  &  0.046  &  0.008   &  0.002  \\
  \hline\hline
\end{tabular*}
\end{table}
Table~\ref{tab:tab1} shows the squared coefficients of this expansion for input beams possessing different OAM values. As seen, the effect of the QP on the radial mode decreases for beams having higher OAM values, so that one can approximately assume that most of the power of the beam is located at the $p=0$ term. If the final detection based on coupling in a single-mode fiber filters only this term, then the presence of the other terms only introduces a certain degree of losses in the system. Hence, within such approximation, the $p$ quantum number plays essentially no role and it can be ignored (except for the Gouy phase issues discussed further below).

Even stronger is the argument one can use if the entire QW simulation takes place in the optical near field. Indeed, at the pupil plane $(\zeta\rightarrow0)$ the expression for the amplitude of HyGG and LG modes simplifies to
\begin{eqnarray}\label{eq:HyGGpupil}
	\hbox{LG}_{p'm'}(\rho,0)&\propto&L_{p}^{|\ell|}(\rho^2)\rho^{|m|}e^{-\rho^2}\\
	\hbox{HyGG}_{pm}(\rho,0)&\propto&\rho^{p+|m|}e^{-\rho^2}.\nonumber
\end{eqnarray}
Combining Eq.\ \ref{eq:qplate} and Eq.\ \ref{eq:HyGGpupil}, it is straightforward to prove that the action of a QP placed at the pupil plane of the beam is given by
\begin{align}\label{eq:qplatepupil}
\widehat Q_{\pi}\mbox{LG}_{0,m}(\rho,0)\ket{L,m}&=-i\,\mbox{LG}_{0,m}(\rho,0)\ket{R,m+1}.
\end{align}
In other words, at the immediate output of the device, the QP ideally results only in the increment of the OAM content, without any alteration of the radial profile. This result remains approximately valid as long as the beam is in the near field, that is for $\zeta\ll1$, except for a region very close to the central singularity and for some associated fringing that occurs outside the singularity. Both these effects can be neglected for $\zeta\ll1$, as the overlap integral of the resulting radial profile with the input Gaussian profile remains close to unity (for example, at $\zeta=0.1$ this overlap is still about 0.93 for a HyGG mode with $m=1$). We exploit this property to minimize any effect due to a possible coupling between the azimuthal and the radial degree of freedom introduced by the QP. The setup was built in order to have all the steps of the QW in the near field of the input photons. To achieve this, we prepared the beam of input photons to have $z_R>10$ m, while the distance between the QW steps was $d\approx10^{-2}z_R$. In the perspective of realizing a QW with high number of steps, a lens system can be used to image the output of each QW unit at the input of the next one; in this way the whole process may virtually occur at the pupil, i.e. at $\zeta=0$, thus effectively canceling all radial-mode effects.

\section{Role of Gouy phases}
Free space propagation of photonic states carrying OAM is characterized by the presence of a phase term, usually referred to as Gouy phase, that evolves along the optical axis. Considering for example LG states of Eq.~\ref{eq:lg}, this phase factor is given by $\exp{\left[-i(2p+|m|+1)\arctan{(z/z_R)}\right]}$, where $z$ is the coordinate on the optical axis with respect to the position of the beam waist. The different phase evolution occurring for different values of $|m|$ could be a significant source of errors in the current implementation of a quantum walk (QW). Let us assume that after the step $n$ in the QW setup the state of the photon is $\ket{\psi}=\sum_m{c_m\ket{m}}$, where for simplicity we consider only modes with $p=0$. When entering the following step, the coefficients $c_m$ will evolve to $c_m'=e^{-i2|m|\arctan{(d/z_R)}}c_m$, where $d$ is the distance between two steps along the propagation axis. At the step $n+1$, coefficients $c_m$ and $c_m'$ lead to different interferences between the OAM paths, altering the features of the QW process. In our implementation we made this effect negligible relying on the condition $d/z_R\ll1$: indeed, as discussed in the previous section, in our setup we had that $z_R>10$ m and $d\simeq10$ cm. An alternative strategy could be based on using a lens system to image each QP on the following one; at image planes all relative Gouy phases vanish. This imaging procedure would thus avoid any effect due to QP contributions to the radial component of the photonic wavefunction, as discussed previously.

\section{Projective OAM measurements on photons}
In order to determine the OAM value of the photons, we have implemented the widely used technique introduced by Mair \emph{et al.} in 2001~\cite{mair:01s}. In this technique the helical phase-front of the optical beam is ``flattened'' by diffraction on a pitch-fork hologram (displayed on a SLM) and the Gaussian component of the beam at the far-field is then selected by a single mode optical fiber. This approach, as shown in Ref.~\cite{hammam:14s}, leads to a biased outcome for the different OAM values, since the coupling efficiency of this projective measurement changes according to OAM of the input beam. For example, the theoretical coupling efficiency for a flattened LG modes to a single Gaussian mode optical fiber with radius $\sigma$ is  
\begin{align}\label{eq:projection}
	\eta_{m}=\frac{2}{\pi \sigma^2}\left|\int_{0}^{\infty}r\,dr\int_{0}^{2\pi}d\phi\,\,{\cal FT}\left[\text{LG}_{0,m}(r,\phi)\,e^{im\phi}\right]\,e^{-\frac{\rho^2}{\sigma^2}}\right|^2,
\end{align}
where ${\cal{FT}}$ is the Fourier transform in the polar coordinates. Obviously, this gives a biased value for different $m$ values, since after being flattened beams have different intensity distributions at the far-field, where the fiber is located. We have taken this effect into account by measuring experimentally the coupling efficiency for different OAM values and then correcting the corresponding measured probabilities.

In the case of two photons, the OAM measurement was carried out in the same way, by previously splitting the beam with a non-polarizing symmetrical beam splitter (BS) and then sending the two output beams on two distinct holograms displayed simultaneously on two portions of the SLM and then coupling both diffracted beams into single-mode fibers.

\section{Quantum walk with different input polarizations}
\begin{figure*}[t]
\centering
\includegraphics[width=16cm]{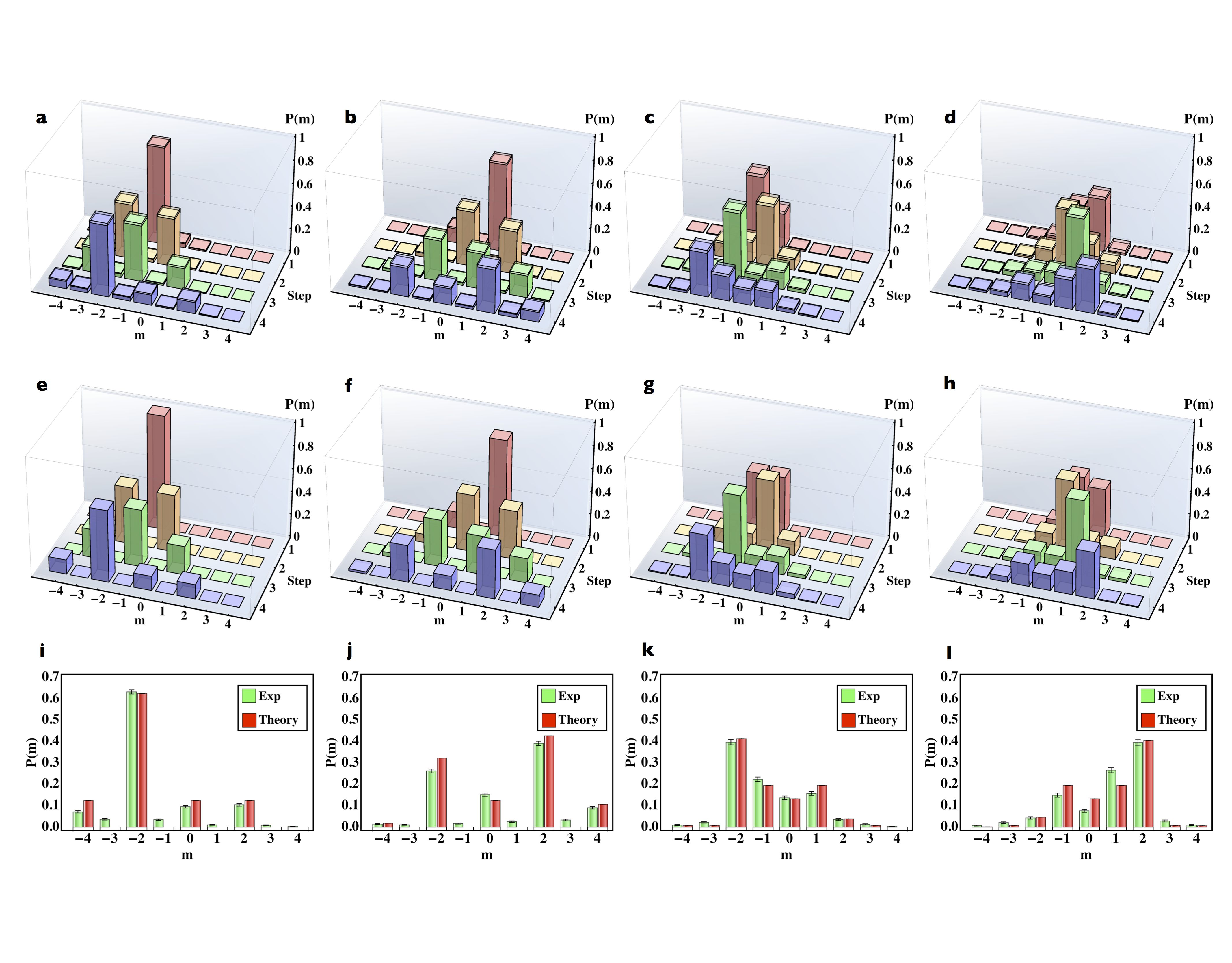}
\caption{Supplementary data for the four-step quantum walk for a single photon, with various input polarization states. a-d) Experimental results, including both intermediate and final probabilities for different OAM states in the evolution (summed over polarizations). The intermediate probabilities at step $n$ are obtained by switching off all QPs that follow that step, that is setting $\delta=0$. Panels a) and b) refer to the standard case with two different input states for the coin subsystem, $(\alpha,\beta)= (1,-1)$ and $1/\sqrt{2} (1/\sqrt 2,1-i/\sqrt2)$, respectively. c) and d) refer to the hybrid case for $\delta=\pi/2$, with the coin subsystem, $(\alpha,\beta)= (1,-1)$ and $1/\sqrt{2} (1-i/\sqrt2,1/\sqrt 2)$, respectively. e-h) Corresponding theoretical predictions. i-l) Comparison of measured and predicted final probabilities. Poissonian statistical uncertainties at plus-or-minus one standard deviation are shown by error bars in panels i-n and as transparent-volumes in panels a-e. The similarities between experimental and predicted OAM distributions are $(89.7\pm0.2)\%$, $(90.9\pm0.6)\%$, $(98.9\pm0.1)\%$ and $(97.0\pm0.4)\%$, respectively. Panels on the same column refer to the same configuration and initial states.}
\label{fig:data1supp}
\end{figure*}

In the case of a single photons, we have carried out measurements with a few other choices of input polarization, besides those already shown in the main article. The results are reported in Fig.\ \ref{fig:data1supp}.

\section{Test of photon correlation inequalities}
Let us consider two photons entering the QW apparatus in fixed states 1 and 2. Here, we use a notation in which the state label at input/output includes both the OAM and the polarization. In our experiment, labels $1,2$ correspond to a vanishing OAM and $L,R$ polarizations. The output states $p$ will denote the combination of the OAM value $m$ and horizontal or vertical linear polarizations $H,V$. The unitary evolution of each photon from these input states to the final states can be described by a matrix $U_{l',l}$, where the first index corresponds to the input state and the second to the output one. Hence, the QW evolution can be described by the following operator transformation law
\begin{equation}
\hat{a}^{\dag}_{l'} \rightarrow \hat{b}^{\dag}_{l'}=\sum_l U_{l',l}\hat{a}^{\dag}_l
\end{equation}

\begin{figure*}[t]
\centering
\includegraphics[width=16cm]{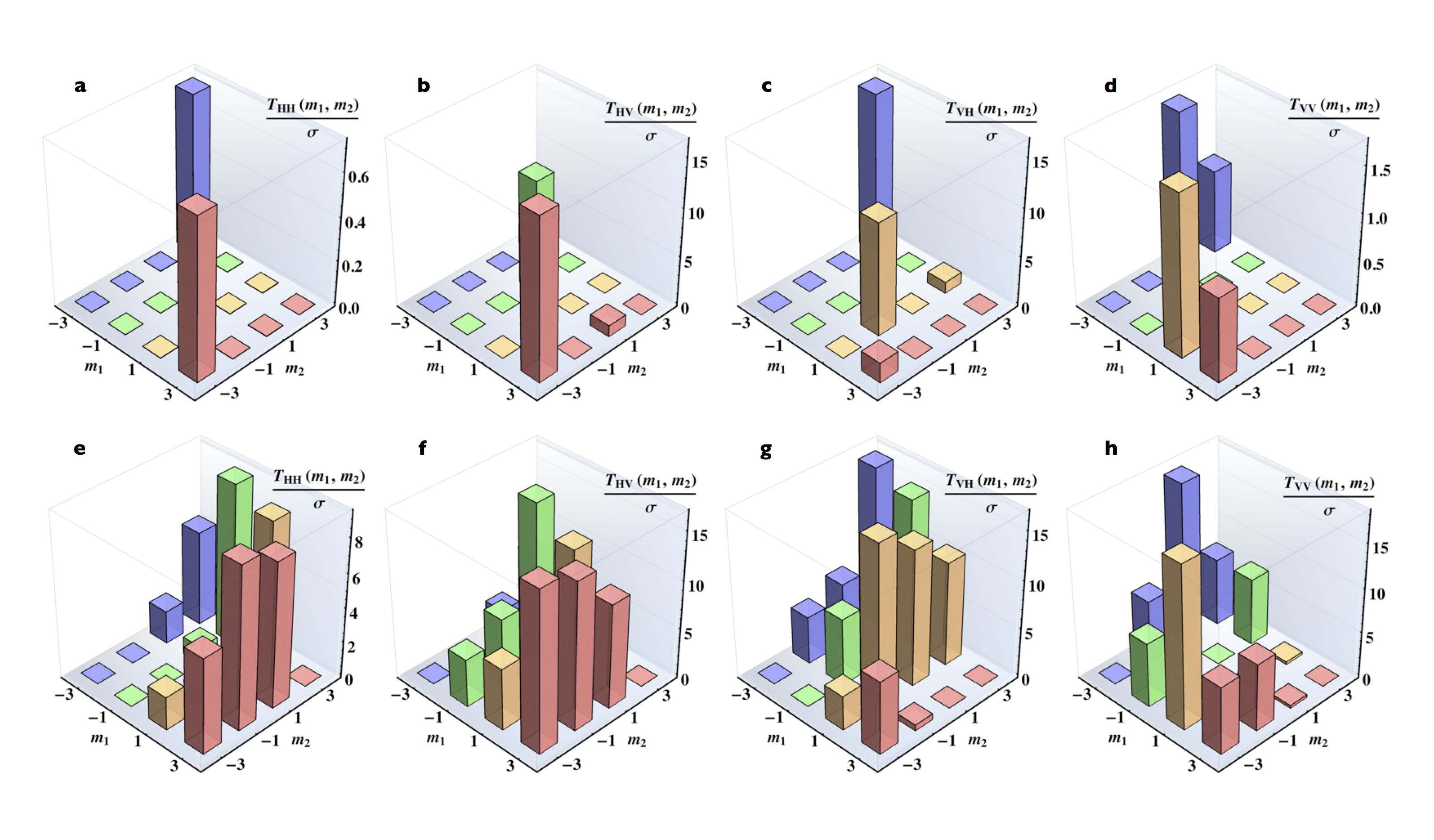}
\caption{Experimental violation of correlation inequalities for two photons which have completed the standard QW ($\delta=\pi$). The data are based on the coincidences after the final beam-splitter. a-d) Violations of the inequalities given in Eq.\ (\protect\ref{classicalinequality}), constraining the correlations that would be obtained for two classical sources, incoherent to each other. Each panel refers to a different pair of measured polarizations for the two photons. These violations prove that our results can only be explained with quantum effects. e-h) Violations of the inequalities given in Eq.\ (\protect\ref{photoninequality}), constraining the correlations obtained for two distinguishable photons. Again, each panel refers to a different pair of polarizations. These violations prove that our photons exhibit two-particle interferences. Only positive values of the $T_{p,q}$ are reported, while negative values which fulfil the inequality are omitted. All violations are given in units of Poissonian standard deviations $\sigma$, as determined from the coincidence counts.}
\label{fig:ineqstandard}
\end{figure*}

\begin{figure*}[t]
\centering
\includegraphics[width=16cm]{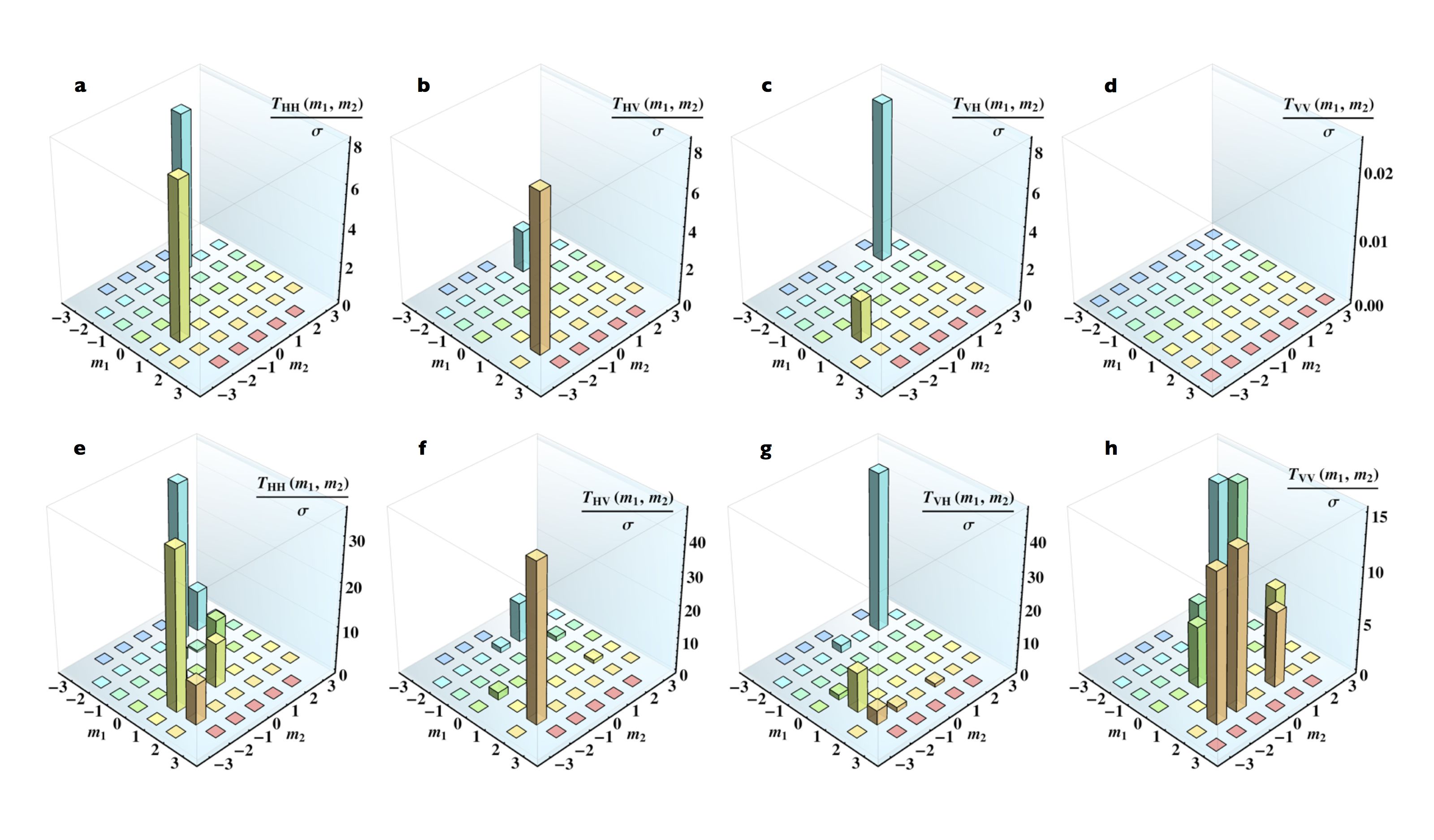}
\caption{Experimental violation of correlation inequalities for two photons which have completed the hybrid QW ($\delta=\pi/2$). The data are based on the coincidences after the final beam-splitter. a-d) Violations of the inequalities given in Eq.\ (\protect\ref{classicalinequality}), constraining the correlations that would be obtained for two classical sources, incoherent to each other. Each panel refers to a different pair of measured polarizations for the two photons. These violations prove that our results can only be explained with quantum effects. e-h) Violations of the inequalities given in Eq.\ (\protect\ref{photoninequality}), constraining the correlations obtained for two distinguishable photons. Again, each panel refers to a different pair of polarizations. These violations prove that our photons exhibit two-particle interferences. Only positive values of the $T_{p,q}$ are reported, while negative values which fulfil the inequality are omitted. All violations are given in units of Poissonian standard deviations $\sigma$, as determined from the coincidence counts.}
\label{fig:ineqhybrid}
\end{figure*}

Let us now discuss the inequalities constraining the measurable photon correlations in two specific reference cases. Our first reference case is that of two independent classical sources (or coherent quantum states with random relative phases) entering modes 1 and 2, in the place of single photons. The following inequality can be then proved to apply to the intensity correlations $\Gamma_{p,q}=\langle \hat{a}^{\dag}_p\hat{a}^{\dag}_q\hat{a}_p\hat{a}_q\rangle$, for any two given QW output modes $p$ and $q$ \cite{bromberg09s,Peru10s}:
\begin{equation}
\frac{1}{3}\sqrt{\Gamma_{p,p}\Gamma_{q,q}}-\Gamma_{p,q}<0.
\end{equation}
In terms of two-photon detection probabilities $\bar{P}_{p,q}=(1+\delta_{p,q})\Gamma_{p,q}$, the same inequality reads
\begin{equation}
\frac{2}{3}\sqrt{\bar{P}_{p,p}\bar{P}_{q,q}}-\bar{P}_{p,q}<0,
\end{equation}
where $\bar{P}_{p,q}$ stands for the probability of having state $\ket{1_p,1_q}$, for $p\neq q$, or state $\ket{2_p}$, for $p=q$, after the QW but before the BS used to split the photons. After the BS, taking into account the photon-splitting probability, the inequality is rewritten as
\begin{equation}
T_{p,q}=\frac{1}{3}\sqrt{P_{p,p}P_{q,q}}-P_{p,q}<0,
\label{classicalinequality}
\end{equation}
where $P_{p,q}$ is now the probability of detecting in coincidence a photon in state $p$ at one (given) BS exit port and the other photon in state $q$ at the other BS exit port.

Our second reference case is that of two single but distinguishable photons entering states 1 and 2. In this case, it is easy to prove a second stronger inequality for the coincidence probabilities. Indeed, in this case one has
\begin{equation}
\bar{P}_{p,q}=|U_{1,p}U_{2,q}|^2+|U_{1,q}U_{2,p}|^2
\end{equation}
for $p\neq q$ and
\begin{equation}
\bar{P}_{p,p}=|U_{1,p}U_{2,p}|^2,
\end{equation}
where $\bar{P}_{p,q}$ now stands for the probability of having one of the two distinguishable photons in state $p$ and the other in $q$ after the QW, before the BS.
The mathematical identity $(|U_{1,p}U_{2,q}|-|U_{1,q}U_{2,p}|)^2>0$ leads directly to the following inequality:
\begin{equation}
2\sqrt{\bar{P}_{p,p}\bar{P}_{q,q}}-\bar{P}_{p,q}<0.
\end{equation}
After the BS, this in turn is equivalent to
\begin{equation}
T_{p,q}=\sqrt{P_{p,p}P_{q,q}}-P_{p,q}<0.
\label{photoninequality}
\end{equation}

The violation of the first inequality (\ref{classicalinequality}) from our coincidence data would prove that the photon correlations cannot be mimicked by intensity correlations of classical sources. Panels (a-d) in Figs.\ \ref{fig:ineqstandard} (standard QW) and \ref{fig:ineqhybrid} (hybrid QW) show the set of violations found in our two-photon experiments, in units of Poissonian standard deviations. In some cases, the experimental violations are larger than 15 standard deviations, proving that the measured correlations are quantum.

The violation of the second inequality (\ref{photoninequality}) from our data proves that the photon correlations are stronger than those allowed for two distinguishable photons, owing to the contribution of two-photon interferences. Although this is already demonstrated in some cases by the violation of the first inequality (as the violation of the first inequality logically implies the violation of the second one), this second inequality is stronger and should be therefore violated in a larger number of cases and with a larger statistical significance (although it requires assuming that there are two and only two photons at input, so that a classical source is excluded a priori). Panels (e-h) in Figs.\ \ref{fig:ineqstandard} (standard QW) and \ref{fig:ineqhybrid} (hybrid QW) show the observed violations. This time, certain measurements violate the inequality by as much as 40 standard deviations, thus proving that two-photon interferences play a very significant role in our experiment.

\end{document}